\documentclass[twocolumn,preprintnumbers,prd,showpacs]{revtex4}
\pdfoutput=1
\usepackage{graphicx}
\usepackage{hyperref}
\usepackage{times}
\usepackage{slashed}

\begin{document}


\title{Discriminating spin through quantum interference}

\author{Matthew R. Buckley$^{1,2}$, Hitoshi Murayama$^{1,2,3}$, William Klemm$^{1}$, and Vikram Rentala$^{1}$}
\affiliation{$^1$ Department of Physics, University of California,
                Berkeley, CA 94720, USA}
\affiliation{$^2$ Theoretical Physics Group, Lawrence Berkeley National Laboratory,
                Berkeley, CA 94720, USA}
\affiliation{$^3$ Institute for the Physics and Mathematics of the Universe, University of Tokyo, 5-1-5 Kashiwa-no-ha, Kashiwa, Japan 277-8568}                
\date{\today}

\begin{abstract}
Many of the proposed solutions to the hierarchy and naturalness problems postulate new `partner' fields to the standard model particles. Determining the spins of these new particles will be critical in distinguishing among the various possible SM extensions, yet proposed methods rely on the underlying models. We propose a new model-independent method for spin measurements which takes advantage of quantum interference among helicity states. We demonstrate that this method will be able to discriminate scalar particles from higher spin states at the ILC, and discuss application to higher spins and possible uses at the LHC.
\end{abstract}
\maketitle

\section{Introduction \label{sec:intro}}
Within the year, the Large Hadron Collider is expected to be up and running, granting us at long last access to the scale of electroweak symmetry breaking and beyond. One of the major puzzles we hope the LHC may provide answers to is the hierarchy problem \cite{Weinberg:1976}\cite{Weinberg:1979bn}\cite{Susskind:1979}\cite{tHooft:1980}: the origin and stability of the orders of magnitude gulf between the Higgs vev at $\sim 300$~GeV and the Planck scale at $\sim 10^{19}$~GeV. Without experimental results, theorists over the years have collected an impressive array of possible solutions to this problem. Arguably, the leading contender is supersymmetry \cite{Wess:1974tw}, but there are many others: extra dimensions \cite{ArkaniHamed:1998rs}\cite{ArkaniHamed:1998nn}\cite{Antoniadis:1998ig}\cite{Antoniadis:1990ew}\cite{Antoniadis:1993jp}\cite{Randall:1999ee}\cite{Randall:1999vf}, technicolor \cite{Weinstein:1973gj}\cite{Weinberg:1979bn}\cite{Susskind:1979}, and little Higgs \cite{ArkaniHamed:2001nc} to name a few. Many of these models also provide a long-lived, weakly coupled particle suitable to be a dark matter candidate. 

In many of these possibilities the immediate experimental signature from the LHC would be the presence of beyond the Standard Model (SM) particles partnered with some or all of the known particles. For example, the minimal supersymmetric standard model (MSSM) doubles the number of particles by adding a new field with the same gauge quantum numbers and Yukawa couplings as in the SM, but spins differing by one half of a unit. Alternatively, the minimal universal extra dimensions (UED) model \cite{Appelquist:2000nn} has compactified extra dimensions which solve the hierarchy problem by `ending physics' at the scale of electroweak symmetry breaking. That is, the Planck scale of the true $4+d$ dimensional theory is not far above a TeV, but appears much larger in $4$D once the compactification occurs. This results in  a tower of KK states, each containing a heavier version of the SM particles with identical quantum numbers, including spin. 

It is well known that, due to the similarities in the particle spectrum and quantum numbers, it may be difficult to distinguish the signatures of MSSM particles from the KK=1 modes of UED at future collider experiments \cite{Cheng:2002ab}\cite{Kong:2006pi}. The existence of the KK=2 modes could serve as a discriminator between supersymmetry and extra dimensions, but their high masses may make them kinematically inaccessible. Even if produced, they typically decay through KK=1 states, and so their presence would only be felt through an increase in the KK=1 production cross section \cite{Battaglia:2005ma}\cite{Burdman:2006gy}. Determining the spin of the new particles will be necessary to confirm the theory underlying any new particles.

There have been several proposals for measuring spin in future collider experiments. The possibilities at a linear collider are far more numerous, due to the control over the center of mass energy in each event. Threshold scans can distinguish scalars from spinors or vector bosons, as the former cross section rises like $\beta^3$ while the latter two are proportional to $\beta$ \cite{Battaglia:2005zf}. However, such a method cannot be used at a hadron collider, and cannot discriminate between spin 1/2 and spin 1. The differential cross section with respect to production angle in $s$-channel pair produced scalars is proportional to $\sin^2 \theta$, while for spinors it is $1+\frac{E^2-m^2}{E^2+m^2}\cos^2\theta$. Model dependence may be present in the form of $t$-channel diagrams, which introduce a forward peak which is similar for both spin statistics \cite{Battaglia:2005zf}. Such diagrams make the production angle measurement of spin more difficult, but may be possible in some cases \cite{Choi:2006mr}.

The polar angular dependence in decays can also be used for spin measurements. However, extracting spin from these measurements assumes knowledge of the final state spins and also requires chiral couplings, introducing a model dependence on the spin measurement \cite{Wang:2006hk}. While this method was originally proposed for the ILC, it was shown that, with sufficiently long decay chains and exploitation of the asymmetry in production of squarks versus antisquarks, supersymmetric spinors could be distinguished from phase space decays at the LHC \cite{Battaglia:2005ma}\cite{Wang:2006hk}\cite{Barr:2004ze}\cite{Barr:2005dz}\cite{Smillie:2005ar}\cite{Alves:2007xt}. Yet this method relies heavily on the underlying models as the entire decay chain must be considered.

Clearly, determination of spin is a problem still requiring novel solutions. In this paper we investigate a model-independent method to determine the spin of new particles at the ILC, first proposed in \cite{MurayamaTalk}. Through interference between the different helicity states in a coherent sum, the cross section of pair produced particles decaying to two body final states develops a non-trivial dependence on the azimuthal angle $\phi$ of the decay. By extracting this dependence, one can determine which helicity states entered into the sum, and thus the spin of the decaying particle. This method is similar to (and was inspired by) the determination of the quark spin measurement at SPEAR \cite{Schwitters:1975dm}. At the end of this paper, we will discuss how this general method may be extended to the LHC.

The paper is organized as follows. In section \ref{sec:calc} we derive the angular dependence of the cross section as a function of particle spin. We then determine appropriate experimental quantities and develop the necessary measurement techniques. In section \ref{sec:scalar} we apply our method to distinguish scalars in pair production at the ILC from production of higher spin states. Spin 1/2 and 1 measurements are considered in section \ref{sec:spinor} and we conclude in section \ref{sec:conc}. Additional calculations are supplied in the Appendices.

\section{Azimuthal Angular dependence \label{sec:calc}} 

To determine the azimuthal dependence of the cross section for pair production followed by decay, we start with a particle of helicity $h$ moving in the $\hat{z}$ direction. When this decays into a two-body final state, the momenta of the daughter particles are confined to a decay plane. If we consider the rotation of this plane about the $\hat{z}$ axis by an angle $\phi$, it is clear that the action of the rotation on matrix element of the decay must be equivalent to the action of the rotation on the parent particle.

Rotations of the particle about the $\hat{z}$ axis introduce a phase $e^{-iJ_z \phi}$, where $J_z$ is the total angular momentum in the $\hat{z}$ direction. However, as the momentum is defined to be in the $\hat{z}$ direction,  the orbital component is zero, and $J_z$ reduces to $h$:
\begin{equation}
J_z = \frac{(\vec{s}+\vec{x}\times\vec{p})\cdot\vec{p}}{|\vec{p}|} = \frac{\vec{s}\cdot\vec{p}}{|\vec{p}|} = h. \label{eq:jz}
\end{equation}
Therefore, the dependence of the decay matrix element ${\cal M}_{decay}$ on $\phi$ must be
\begin{equation}
{\cal M}_{decay}(h, \phi) = e^{ih\phi}{\cal M}_{decay}(h,\phi=0). \label{eq:mphi}
\end{equation}
Were we to produce particles in only one helicity state, then the total cross section (proportional to the square of ${\cal M}_{prod.}{\cal M}_{decay}$) would be independent of $\phi$. However, if more than one helicity states are produced and then decay, the total cross section is proportional to the coherent sum squared:
\begin{equation}
\sigma \propto \left|\sum_h  {\cal M}_{prod.}(h)e^{ih\phi}{\cal M}_{decay}(h,\phi=0) \right|^2. \label{eq:corsum}
\end{equation}
This expression is true only within the validity of the narrow-width approximation. However, for `weakly-coupled' physics, such an approximation is justified \cite{Berdine:2007uv}.

As a result of this interference among the various helicity states, the cross section develops a $\cos(n \phi)$ dependence, where $n$ is an integer running from zero to twice the largest value of $h$ for the decaying particle. That is to say, the $\phi$ dependences for a scalar, spinor, and massive vector boson can be written as
\begin{eqnarray}
\sigma(s=0) & = & A_0 \label{eq:scalar} \\
\sigma(s=1/2) & = & A_0+A_1 \cos(\phi) \label{eq:spinor} \\
\sigma(s=1) & = & A_0+A_1\cos(\phi)+A_2 \cos(2\phi) \label{eq:vector}
\end{eqnarray}
where the $A_i$ are not functions of $\phi$ (though they are non-trivial functions of the other kinematics of the problem). The exact forms of the cross section must be worked out from the standard rules of constructing matrix elements from Feynman diagrams, in which case the $\phi$ dependence will become apparent. However, from this general argument the relationship between spin and $\phi$ dependence is made clear.

\begin{figure}[h]
\centering
\includegraphics[width=0.5\columnwidth]{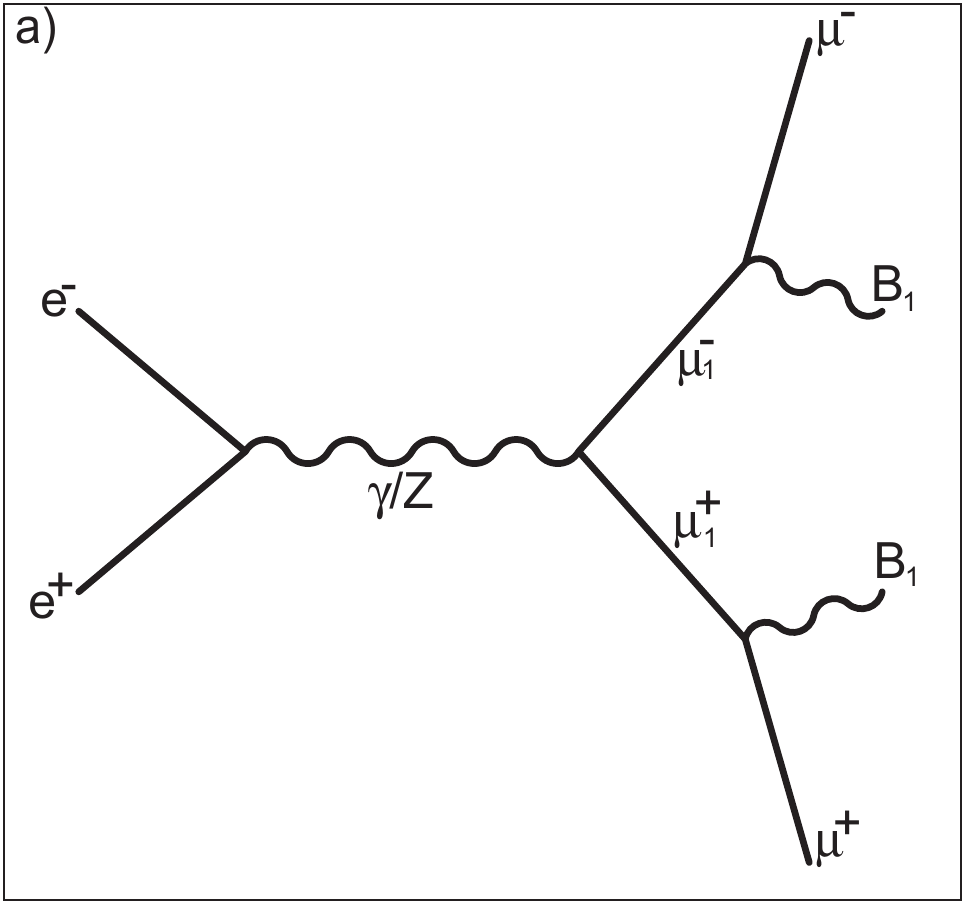}\includegraphics[width=0.5\columnwidth]{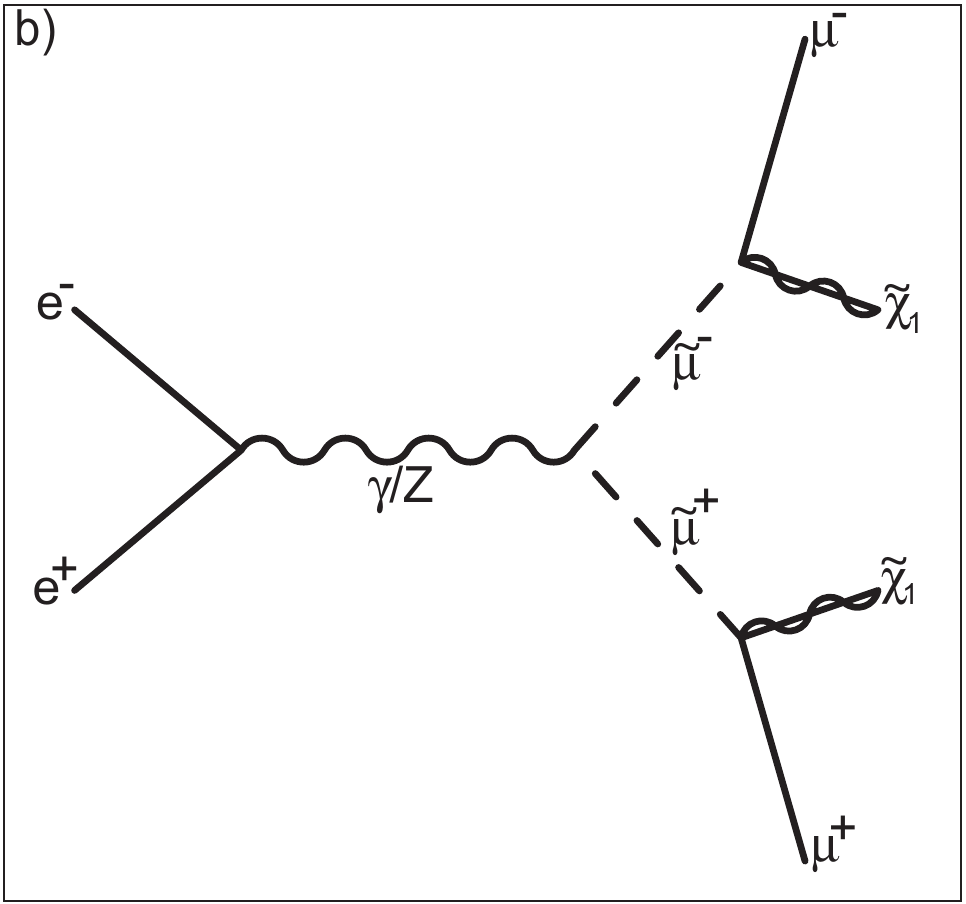}

\caption{a) Pair production of KK$=1$ muons in universal extra dimensions decaying to opposite sign muons and missing energy in the form of two $B_1$ gauge bosons (the LKP). 
b) Pair production of smuons in supersymmetry decaying to opposite sign muons and the lightest neutralinos as LSP missing energy \label{fig:feyn1}}
\end{figure}

To exploit this $\phi$ dependence, we consider pair production of particles from $e^+e^-$ at the future ILC. As motivated by solutions to the dark matter problem, we expect the production of beyond the SM particles to cascade down to some weakly coupled particle which will escape the detector. Such WIMPs are present in the supersymmetric spectrum as the lightest supersymmetric particle (LSP), typically the lightest neutralino; or in universal extra dimensions as the lightest Kaluza-Klein odd particle (LKP), typically the $B_1$. Examples of such event topologies in the UED and SUSY models are shown in Fig.~\ref{fig:feyn1}. However, our methods do not rely on such specific models.

Measuring the azimuthal dependence of the cross section requires that we are able to reconstruct the momentum of the parent particle. For simplicity, we specialize to cases where the pair-produced particles each decay to a charged lepton and missing energy, in which case the events of interest consist of $\ell^\pm \ell^\mp \slashed{E}$. While we risk losing some model independence at this stage, such signatures are fairly generic in many extensions to the SM.

\begin{figure}[h]
\centering
\includegraphics[width=\columnwidth]{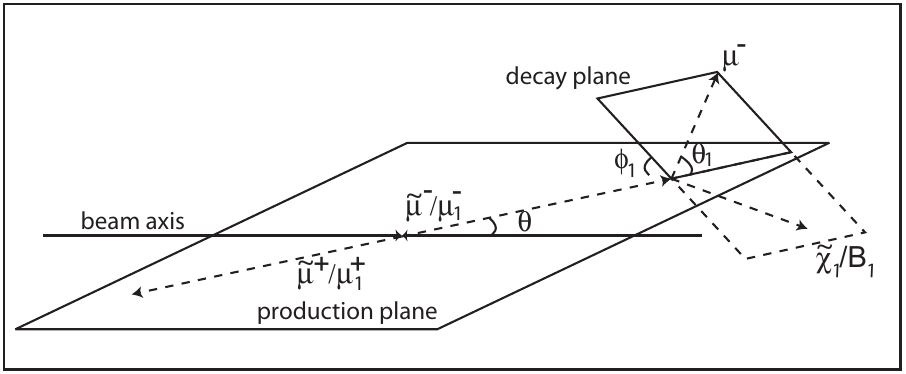}

\caption{The pair produced $\tilde{\mu}$ or $\mu_1$ in the lab frame. The beam axis is defined as the $z$ axis, with the production angle $\theta$ in the $x-z$ plane. The $\hat{z}$ axis is defined to point along the production axis. The decay angle $\phi_1$ is invariant to boosts along $\hat{z}$, and so may be defined in either the lab frame or the frame of the decaying particle. The angle $\theta_1$ is defined in the rest frame of $\tilde{\mu}^-/\mu_1^-$. Decay angles $\theta_2$ and $\phi_2$ (not shown) are defined equivalently for the $\tilde{\mu}^+/\mu_1^+$. \label{fig:angles}}
\end{figure}

Let the pair produced particles who's spin we wish to measure ($\mu_1$ in Fig.~\ref{fig:feyn1}a or $\tilde{\mu}$ in Fig.~\ref{fig:feyn1}b) have 4-momenta $p_A$ and $p_B$ and mass $M$ . These particles each decay to visible (effectively massless) leptons and a weakly coupled particle with mass $m$ (the $\tilde{\chi}_1^0$ or $B_1$ in Fig.~\ref{fig:feyn1}). We label the visible lepton momentum $p_1$ and $p_2$, and assume that the particles running in each leg are identical. The production angle $\theta$ and decay angles $\theta_i$ and $\phi_i$ ($i=1$ for the decay of $A$ and $i=2$ for the decay of $B$) defined relative to the production plane are shown in Fig.~\ref{fig:angles}.

At the ILC, assuming knowledge of the masses $M$ and $m$, it is possible to completely reconstruct the 4-momenta $p_A$ and $p_B$ (and thus the angles $\phi_1$ and $\phi_2$) up to a two-fold ambiguity \cite{Tsukamoto:1993gt}\cite{Cheng:2007xv}. We note that there are $4$ unknown values for both of the missing particles in the decay, for a total of 8 unknowns. There are 4 measured values of the total missing $4$-momentum $\slashed{p}$; and for each massive particle in the diagram there is a mass relation, for a total of 4 constraints. Therefore, one would expect this event to be completely reconstructible. When solving the mass relations however, one finds an ambiguity in sign when taking a square root, leading to the two-fold ambiguity in the reconstructed momentum. For the details of the reconstruction, see Appendix \ref{sec:recon}.

With less than perfect knowledge of the masses, muon momenta and center of mass energy (from beamstrahlung \cite{Wilson:1986by} and initial state radiation), the true solution will not be reconstructed perfectly. At the ILC, masses of lepton and gaugino partners are expected to be measured to one part per mille \cite{Freitas:2002gh}\cite{Feng:2001ce}, the tracking resolution as good as $\sim 5 \times 10^{-5} (p_T/\mbox{GeV})$ \cite{Behnke:2001qq}, and beamstrahlung/ISR should be a few percent \cite{Murayama:1996ec}. As such, we expect that the errors introduced in $\phi$ from these effects will be minimal.

As we have two solutions for the momenta $p_A$ and $p_B$, this leads to two solutions each for $\phi_1$ and $\phi_2$. The extracted signal in the azimuthal distribution is therefore obtained in the combination of the true and false solutions and compare to the expected values given in Eqs.~(\ref{eq:scalar}), (\ref{eq:spinor}), and (\ref{eq:vector}). In particular, a particle of spin $n/2$ should have $A_i =0$ for all $i>n$. 

\section{Scalars vs. Spinors \label{sec:scalar}}

To demonstrate the utility of this method of spin determination, we consider the pair production of scalar right-handed smuons in supersymmetry which decay to muons and LSP $\tilde{\chi}_1^0$s ($e^-e^+ \to \tilde{\mu}^-_{R}\tilde{\mu}^+_{R} \to \mu^-\mu^+\tilde{\chi}_1^0\tilde{\chi}_1^0$).  We compare the azimuthal distributions of $\phi_1$ and $\phi_2$ in this scenario to that in the pair production of $\mu_{1R}$s decaying to muons and LKP $B_1$s in a UED model ($e^-e^+ \to \mu^-_{1R}\mu^+_{1R} \to \mu^-\mu^+B_1B_1$). The Feynman diagrams for these processes are shown in Fig.~\ref{fig:feyn1}. Analytic formulae for the production and decay cross sections for both models are presented in Appendix \ref{sec:cross}. We stress that SUSY and UED are chosen only as benchmark models with differing spins and similar final states, the method used to determine spin can in principle work equally well for any other scenario.

Representative spectra are required for both supersymmetry and universal extra dimensions. In addition, we wish to avoid any possible model-specific effects on the azimuthal distributions arising from different choices of spectra for supersymmetry and extra dimensions. Therefore, as the masses of the $\tilde{\mu}/\mu_1$ and $\tilde{\chi}_1^0/B_1$ are assumed to be known, we perform our analysis twice for each model. In the first case we assign the masses of the $\mu$ and $B$ partners as per a SUSY spectrum, and then repeat the processes with the UED case.

As a representative supersymmetry point, we chose SPS 3 \cite{Ghodbane:2002kg}\cite{Allanach:2002nj} in mSUGRA parameter space, which has $m_0 = 90~\mbox{GeV}$, $m_{1/2} = 400~\mbox{GeV}$, $A_0=0$, $\tan\beta = 10$ and a positive $\mu$ parameter. Universal extra dimensions is represented by the minimal version (MUED) \cite{Cheng:2002iz}, which requires only four parameters: the number and radius $R$ of the extra dimensions, the scale $\Lambda$ to set flavor-universal boundary terms equal to zero, and the Higgs mass. We chose one extra dimension with $R^{-1} = 300~\mbox{GeV}$, $\Lambda = 20R^{-1}$ and a Higgs mass of $120~\mbox{GeV}$. The resulting particle spectra at the TeV scale are shown in Table~\ref{tab:spectra}. 

\begin{table}
\centering
\begin{tabular}{|c|c|c|}
\hline
 & SPS 3 & MUED \\ \hline \hline
 $\tilde{\chi}_1^0/B_1$ & $161$~GeV & $302$~GeV\\ \hline
 $\tilde{\ell}_R/\ell_{1R}$ & $181$~GeV & $304$~GeV \\ \hline
 $\tilde{\ell}_L/\ell_{1L}$ & $289$~GeV & $309$~GeV \\ \hline
 $\tilde{\chi}_1^\pm/W_1^\pm$ & 306~GeV & $327$~GeV  \\ \hline
 $\tilde{\nu}_\ell/\nu_{1\ell}$ & 276~GeV & $309$~GeV  \\ \hline
\end{tabular}

\caption{Relevant particle spectra for the mSUGRA parameter point SPS 3 and the minimal universal extra dimension parameters $R^{-1}=300~\mbox{GeV}$, $\Lambda=20R^{-1}$ and $m_H = 120~\mbox{GeV}$. The MUED spectrum was derived using the MUED package \cite{mued} for CalcHEP   \cite{Pukhov:2004ca}. $\ell$ here refers to the light charged leptons: electrons or muons.  \label{tab:spectra}}
\end{table}

Backgrounds consist of the standard model production of $W^-W^+$ pair production with leptonic decays to muons and neutrinos, $ZZ$ production with decays to muons and neutrinos, and model-background of $\tilde{\chi}_1^+\tilde{\chi}_1^-/W_1^+W_1^-$ production decaying to muons and $\tilde{\nu}/\nu_1$. While kinematic cuts on the invariant mass of the muon pairs can greatly reduce the SM background, more efficient cuts can be obtained by requiring successful reconstruction of the $\tilde{\mu}_R/\mu_{1R}$ momentum as outlined in Appendix~\ref{sec:recon}.

The reconstruction algorithm assumes that the masses of the produced and escaping particles are known. By assuming that the signature of $\mu^-\mu^+\slashed{E}$ arises from pair production of $\tilde{\mu}_{R}$ (or $\mu_{1R}$) decaying to LSP or LKP, all other events with the same signature but different particle masses develop inconsistencies in their reconstruction. That is, the visible momenta are not compatible with the pair production of particles with masses other than that for the $\tilde{\mu}_R/\mu_{1R}$ decaying into particles with masses other than that of the LSP/LKP. In practice, the parameter $y$ defined in Eq.~(\ref{eq:y}) becomes imaginary. 

With perfect knowledge of masses and muon momentum, requiring reconstruction to succeed cuts nearly all of the background events. Once detector smearing and mass measurement errors are included, it is inevitable that some background will survive the reconstruction cut. Again, with the small errors in mass measurements available at the ILC we do not expect large backgrounds to pollute the data set. 

\begin{figure}
\centering
\includegraphics[width=0.5\columnwidth]{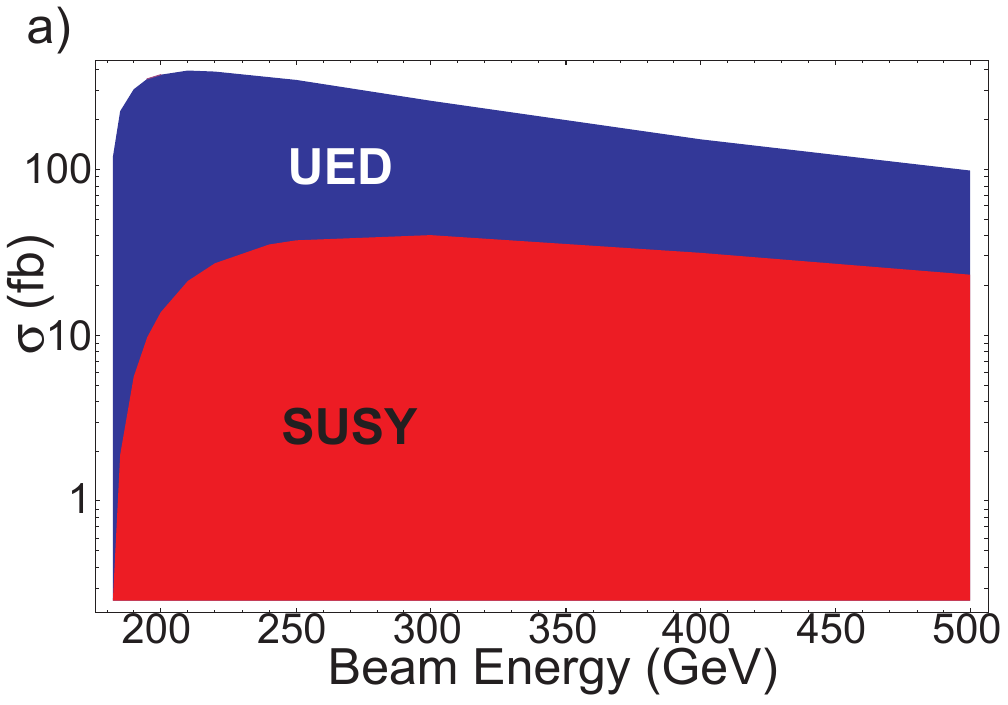}\includegraphics[width=0.5\columnwidth]{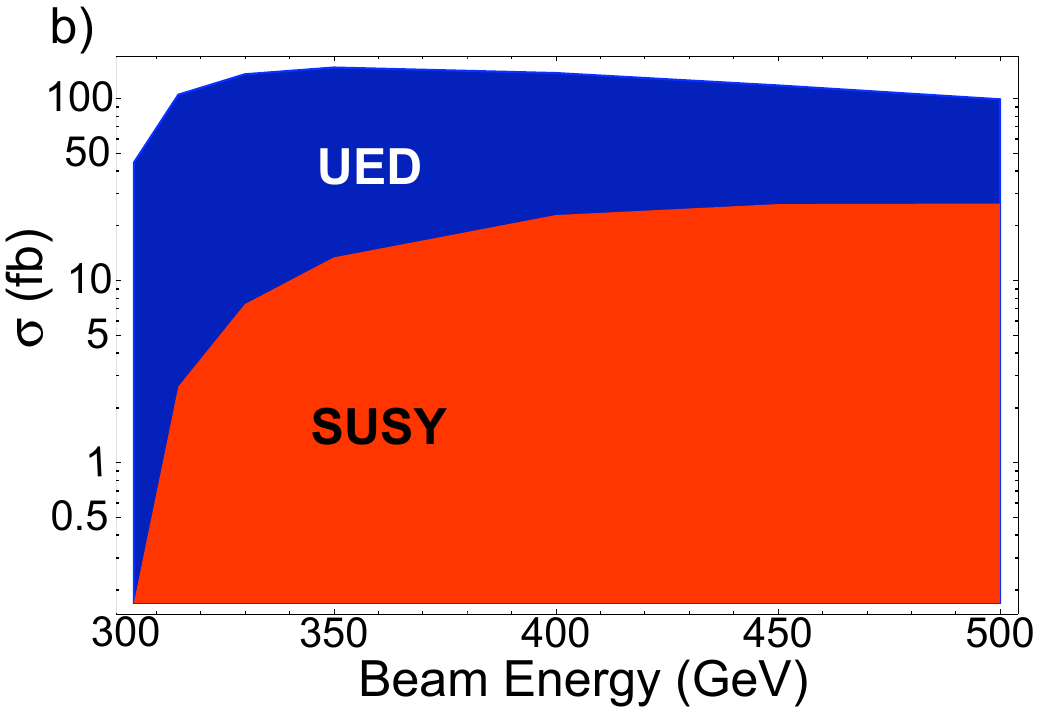}

\caption{Cross sections times branching ratios as a function of the beam energy for the UED process $e^-e^+ \to \mu^-_{1R}\mu^+_{1R} \to \mu^-\mu^+B_1B_1$ and the SUSY process $e^-e^+ \to \tilde{\mu}^-_{R}\tilde{\mu}^+_{R} \to \mu^-\mu^+\tilde{\chi}_1^0\tilde{\chi}_1^0$. Figure a) uses the SPS 3 spectrum, while b) uses the MUED spectrum (see Table~\ref{tab:spectra}). \label{fig:sigma}}
\end{figure}

The total center of mass energy at the ILC is expected to reach up to $1~\mbox{TeV}$, and an integrated luminosity of $500~\mbox{fb}^{-1}$ is not unrealistic. For the mass spectra chosen, the resulting cross sections times branching ratios are shown in Fig.~\ref{fig:sigma} for $\sqrt{s}$ running from threshold up to $1$~TeV. As a result, we expect several thousand to several hundreds of thousands of events available. 

To simulate the effects of necessary cuts due to the geometry of the detector, we place cuts on the pseudo-rapidity $\eta$. We require $\eta$ to be less than $2.5$ for both visible muons, as otherwise the leptons would vanish unseen down the beam. Also, if the missing momentum also points down the beam pipe we cannot be sure that the missing energy is truly in the form of WIMPs and not merely unobserved SM particles, so we cut on $\eta$ for missing $\vec{p}_T$ as well.

Using HELAS \cite{Murayama:1992gi} the production and decay matrix elements were calculated at tree level for each helicity state. Using the narrow width approximation, the cross sections as a function of $\theta$, $\phi_1$, $\theta_1$, $\phi_2$, and $\theta_2$ were obtained. BASES \cite{Kawabata:1985yt}, an adaptive Monte Carlo, program was used to integrate over the other kinematic angles to determine the differential cross sections with respect to $\phi_1$ and $\phi_2$. As both decaying particles have the same spin statistics, the differential distributions are the same for both $\phi$s and so, to increase statistics, the distributions for $\phi_1$ and $\phi_2$ were added. 

\begin{figure}
\centering
\includegraphics[width=\columnwidth]{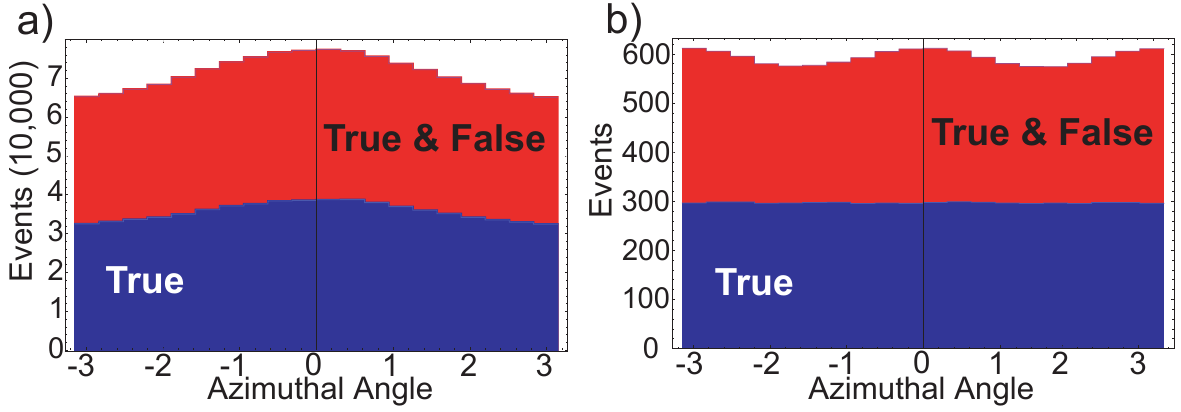}

\caption{Histograms of number of events per azimuthal angle $\phi$ for both the true solution to the reconstruction algorithm and the combined true and false distribution. The center of mass energy is $\sqrt{s}=370$~GeV and the luminosity if $500~\mbox{fb}^{-1}$. Figure a) shows the UED distribution for $e^-e^+ \to \mu^-_{1R}\mu^+_{1R} \to \mu^-\mu^+B_1B_1$ while b) is the SUSY distribution for $e^-e^+ \to \tilde{\mu}^-_{R}\tilde{\mu}^+_{R} \to \mu^-\mu^+\tilde{\chi}_1^0\tilde{\chi}_1^0$. \label{fig:dist}}
\end{figure}

Representative distributions for scalar and spinors (including rapidity cuts) are shown in Fig.~\ref{fig:dist}. As can be clearly seen in Fig.~\ref{fig:dist}a, both the true and false UED distributions have clear $\cos\phi$ dependence, as expected from spinor decay (Eq.~(\ref{eq:spinor})). The true distributions for the scalar SUSY decay in Fig.~\ref{fig:dist}b is flat, as expected from Eq.~(\ref{eq:scalar}). It is therefore apparent even at this level of analysis that the $\phi$ dependence of the distribution contains the spin information necessary for our method.

Considering the combined true and false distribution in Fig.~\ref{fig:dist}, a systematic issue for our method becomes readily apparent. An unexpected $\cos2\phi$ dependence develops due to the false distribution and rapidity cuts, a situation we regard as an indication of practical limitations to our method, not a fundamental flaw. Whereas the $\cos 2\phi$ dependence may be unimportant for the discrimination of scalar versus higher spin states, it will become important in distinguishing spinor from vectors (section \ref{sec:spinor}). Though harder to see by eye, the UED distribution also develops a $\cos2\phi$ dependence in the false solution. As such, we fit not to $A_0+A_1\cos\phi$ but rather to $A_0+A_1\cos\phi+A_2\cos2\phi$.

The overall scaling of the $A_i$ parameters in Eqs.~(\ref{eq:scalar}), (\ref{eq:spinor}), and (\ref{eq:vector}) depends on the total number of events, which is a function of the total cross section. To remove this model dependent effect, the parameter of interest in spin determination is not $A_1$, but $A_1/A_0$, which is independent of the scaling due to total cross section.

\begin{figure}
\centering
\includegraphics[width=\columnwidth]{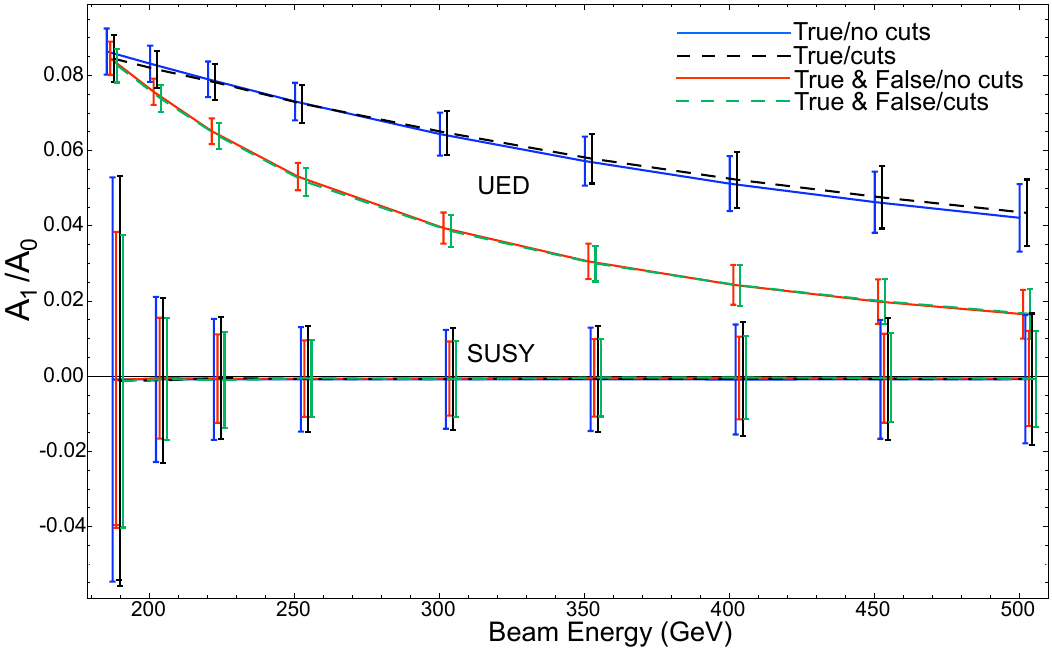}

\includegraphics[width=\columnwidth]{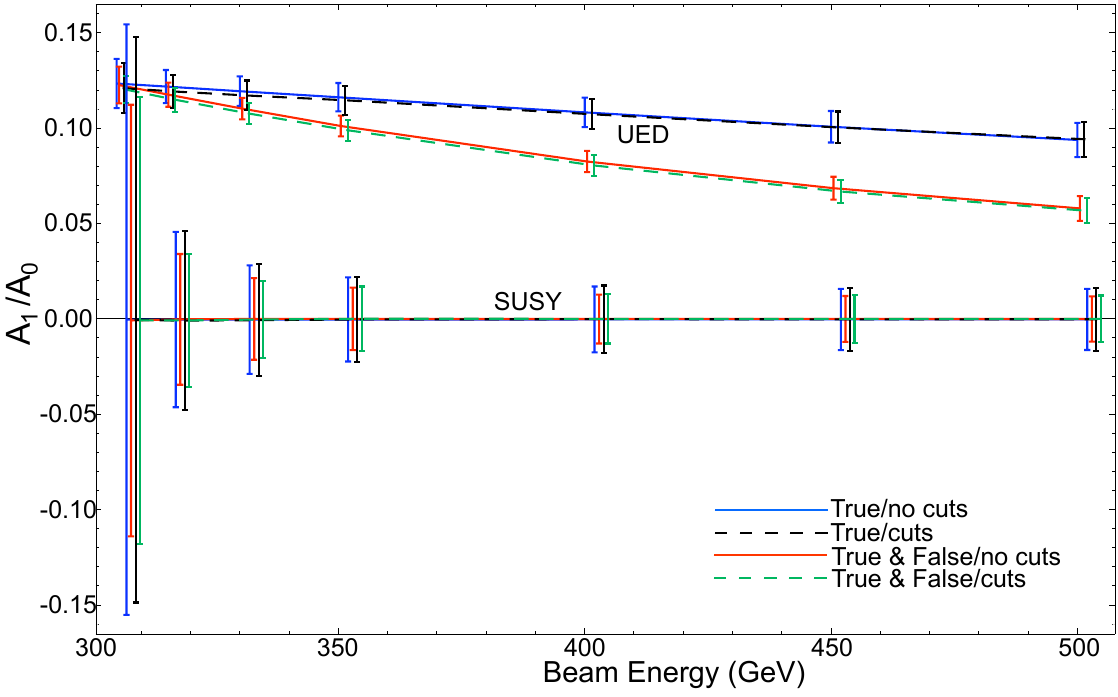}

\caption{Top: Ratio $A_1/A_0$ for mSUGRA parameter point SPS3 as a function of energy for both scalar (SUSY) and spinor (UED) pair production with $500~\mbox{fb}^{-1}$ luminosity.. Error bars correspond to $95\%$ exclusion region. Blue lines correspond to true solution only with no rapidity cuts, black dashed lines are true solutions with rapidity cuts, red lines to true and false solutions without cuts, and green are true and false solutions with cuts. Bottom: Ratio $A_1/A_0$ for MUED parameters as in Table~\ref{tab:spectra} for both scalar (SUSY) and spinor (UED) production. Color labeling identical to the above. \label{fig:a1a0}}
\end{figure}

Using the least squares method the generated distributions were fit to $A_0+A_1\cos\phi+A_2\cos2\phi$. The errors for each parameter $A_i$ were obtained after marginalizing over the other two parameters. The ratio $A_1/A_0$ for the scalar $\tilde{\mu}_R$ and spinor $\mu_{1R}$ are shown in Fig.~\ref{fig:a1a0}. As can be seen, for both the SPS3 and MUED spectra the values of $A_1/A_0$ for $\tilde{\mu}_R$ are consistent with zero for all energies and for both the true and false distributions. For the spinor $\mu_{1R}$, the ratio is manifestly non-zero, allowing us to distinguish scalars from higher spin states.

Several aspects of Fig.~\ref{fig:a1a0} require closer examination. The large error bars for the supersymmetric particles in both spectra are due to the relatively poor statistics compared to the pair production of the spinor KK modes in universal extra dimensions. This is especially apparent near threshold. For the spinor particles we also note that, near threshold, the signal is on the order of $10\%$, and decreases towards zero at progressively higher energies. This decrease can be readily explained as follows: far from threshold, the mass of the pair produced particles becomes less relevant, and so their spins become more correlated due to chirality conservation. To determine the distribution of $\phi_1$ ($\phi_2$), we integrate over all other angles in the problem, including $\phi_2$ ($\phi_1$). Due to the correlation of spins in this energy regime, this integration results in decoherence of the sum of matrix elements. That is, rather than considering $\left|\sum_h {\cal M}(h)\right|^2$, at high energies the cross section becomes proportional to $\sum_h\left|{\cal M}(h)\right|^2$, which has no azimuthal angle dependence due to the lack of interference between terms.

Finally, in considering the distribution of true solutions versus that of the combined solutions, we note that for the spinor case the signal is less once the false solutions are added in. At low energies the difference between the two is comparatively small, but grows as we move away from threshold. This agrees well with the naive intuition that the false distribution should be flat in $\phi_1$ and $\phi_2$; however we stress that at higher frequencies such intuition fails us and the flat distribution may develop non-trivial $\cos 2\phi$ dependences.

\begin{figure}
\centering
\includegraphics[width=\columnwidth]{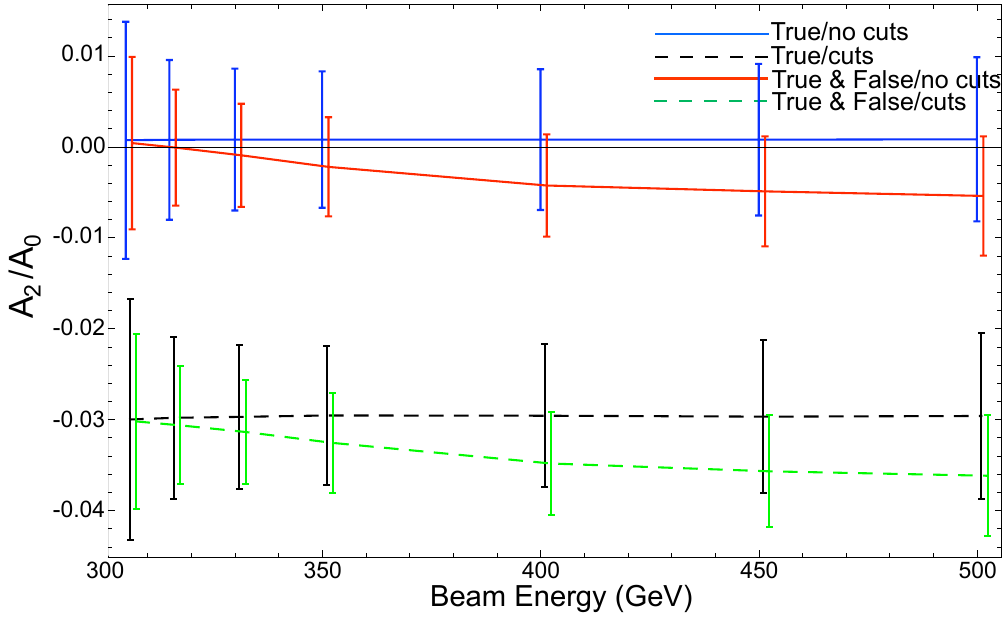}

\includegraphics[width=\columnwidth]{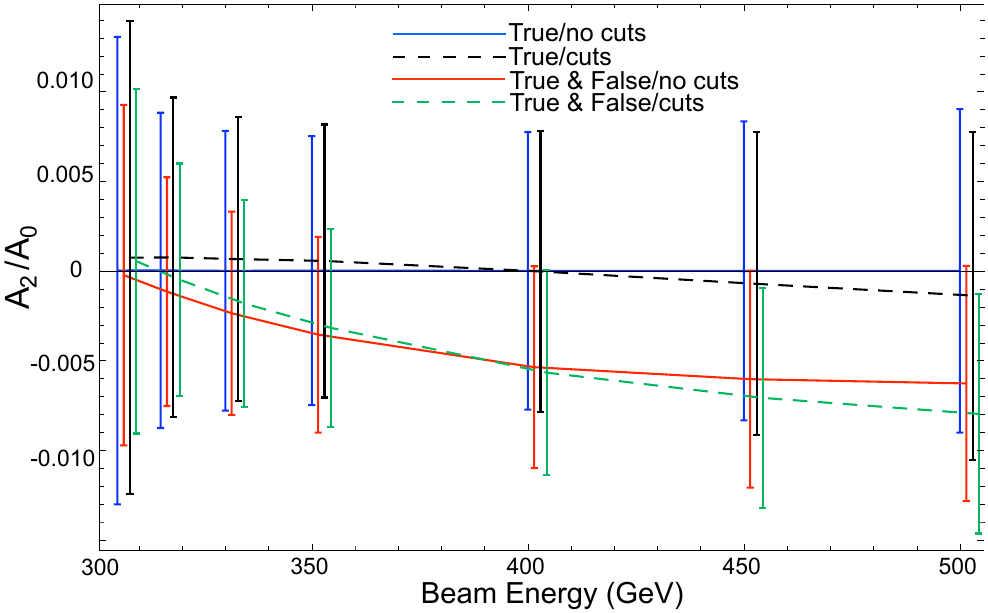}

\caption{Top: Ratio $A_2/A_0$ for mSUGRA parameter point SPS3 as a function of energy for spinor (UED) pair production with $500~\mbox{fb}^{-1}$ luminosity. Error bars correspond to $95\%$ exclusion region. Blue lines correspond to true solution only with no rapidity cuts, black dashed lines are true solutions with rapidity cuts, red lines to true and false solutions without cuts, and green are true and false solutions with cuts. Bottom: Ratio $A_2/A_0$ for SPS3 parameters for spinor (UED) production after correcting for effects of false distribution and cuts on a flat distribution. Color labeling identical to the above. \label{fig:a2a0}}
\end{figure}

To demonstrate this effect we plot in Fig.~\ref{fig:a2a0} the ratio $A_2/A_0$ for the decay of spinor $\mu_{1R}$ (using SPS3 parameters). As can be seen in the top plot, the true solution without cuts has a coefficient of zero for the $\cos 2\phi$ term, as predicted by Eq.~(\ref{eq:spinor}) for spinor decay. However once cuts and the false solutions are added a non-zero value is generated. Clearly, this can cause confusion between a spin-1/2 particle and a vector or higher spin state. 

To attempt to correct for this effect we generate events in which the particles decay according to phase space. This flat distribution is reconstructed using the method outlined in Appendix~\ref{sec:recon} and rapidity cuts are applied just as in the SUSY and UED cases. As a result, the flat distributions also develop a $\cos2\phi$ dependence. The resulting values for $A_2/A_0$ using only true solutions (with and without cuts) and then both solutions (with and without cuts) are subtracted from the appropriate spinor ratios to isolate the spin-dependent effect. The resulting corrected $A_2/A_0$ values are displayed in the lower plot in Fig.~\ref{fig:a2a0}. As can be seen, the flat distribution corrects the $\cos2\phi$ contribution due to cuts but does not remove the false distribution's effect, leaving a $\sim 0.5\%$ spurious signal at high energies. For reasons we do not yet fully appreciate, at low energies the false distribution's effects are minimal, allowing for the possibility of accurate spin measurements. However, it is exactly in this regime that statistics are poor due to the proximity of the threshold.

\section{Spinor vs. Vector \label{sec:spinor}}

Due to the large $A_1/A_0$ signal for non-scalars (on the order of $10\%$) and minimal effect of rapidity cuts and false distributions on this ratio, the ILC should have little difficulty discerning that a particle is spin-0. However for higher spins the cuts and false solutions introduce potentially dangerous higher frequency contributions, as has been demonstrated. 

\begin{figure}[h]
\centering
\includegraphics[width=0.5\columnwidth]{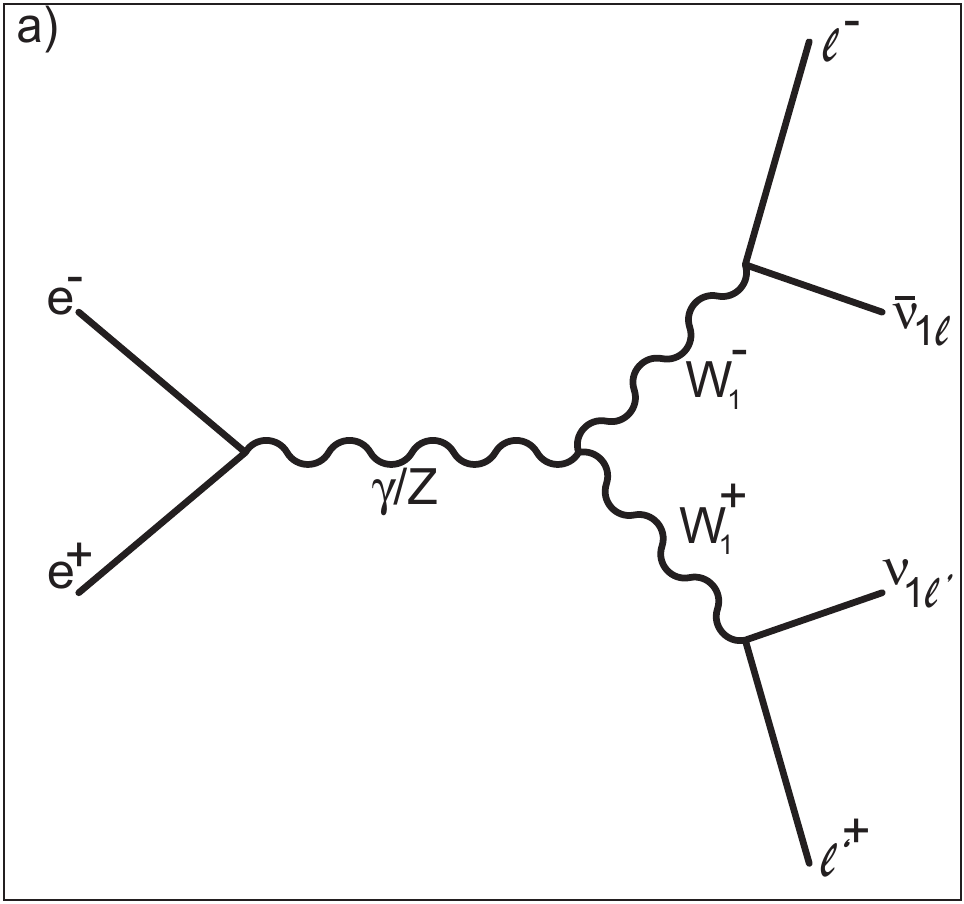}\includegraphics[width=0.5\columnwidth]{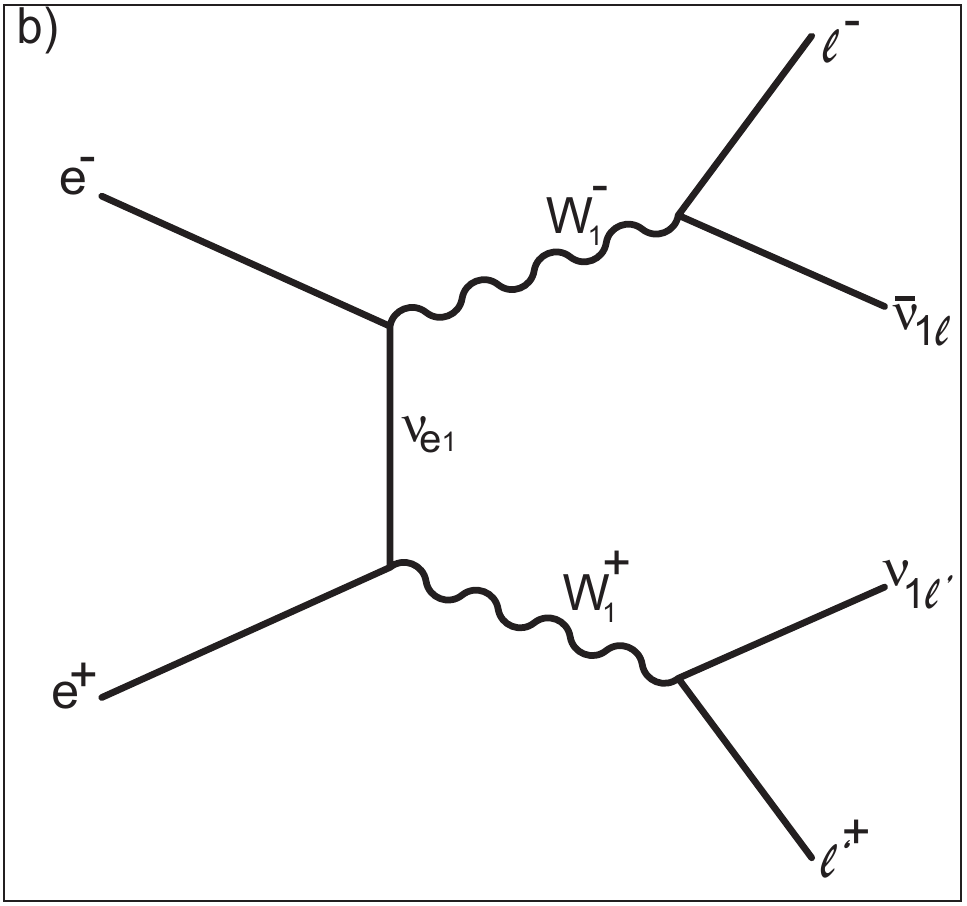}
\includegraphics[width=0.5\columnwidth]{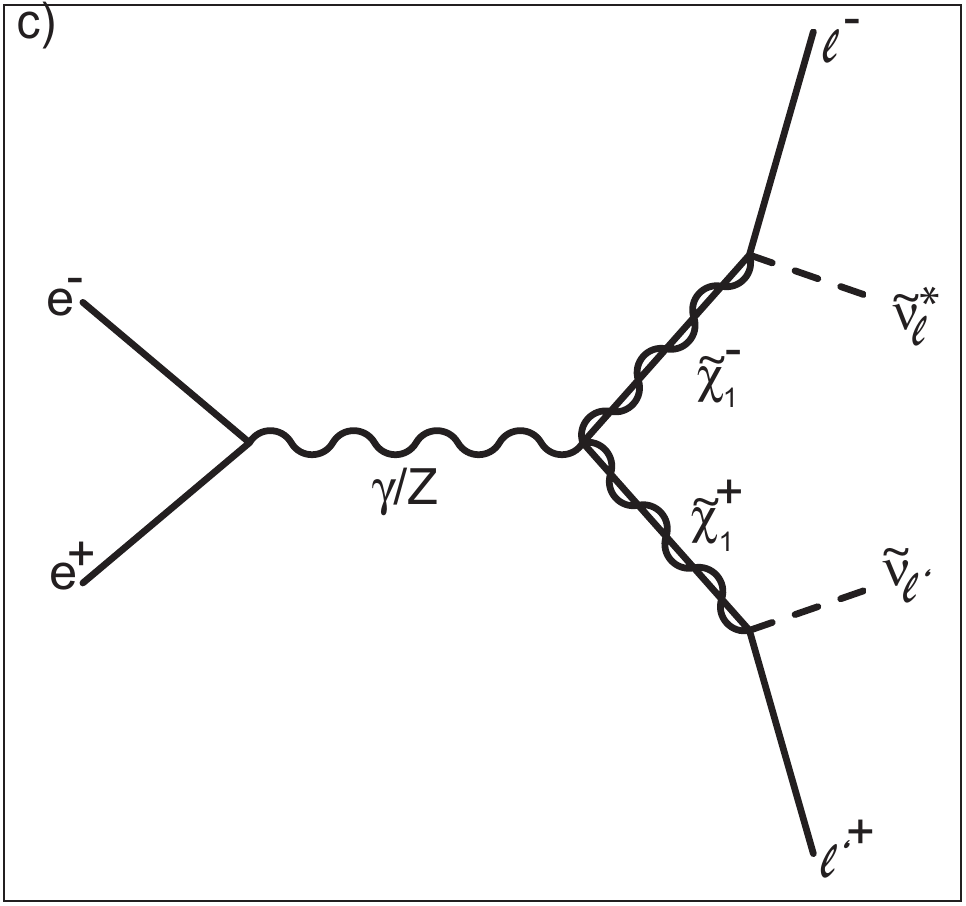}\includegraphics[width=0.5\columnwidth]{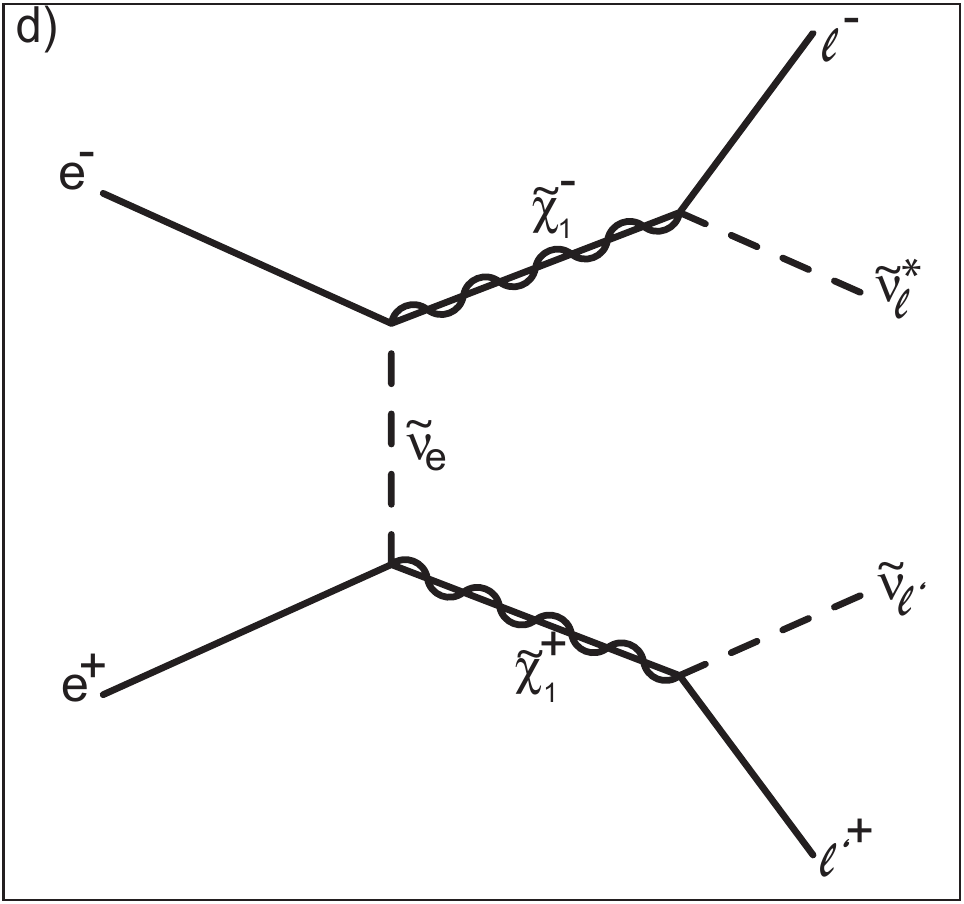}

\caption{a) $s$-channel and b) $t$-channel pair production of KK$=1$ W bosons in universal extra dimensions decaying to opposite sign leptons and missing energy in the form of two $\nu_1$s. 
c) $s$-channel and d) $t$-channel pair production of charginos $\tilde{\chi}_1^\pm$ in supersymmetry decaying to opposite sign leptons and sneutrino missing energy \label{fig:feyn2}}
\end{figure}

As a result, the question still remains whether this method may be practically applied to discriminate spinors from vectors in general cases. We therefore consider a case of pair production of massive vector bosons in UED contrasted with spinor production in SUSY. In particular, we consider $e_L^-e_L^+ \to W_1^-W_1^+\to \ell^- \ell'^+ \bar{\nu}_{1\ell}\nu_{1\ell'}$ in universal extra dimensions and $e_L^-e_L^+ \to \tilde{\chi}_1^-\tilde{\chi}_1^+ \to \ell^- \ell'^+\tilde{\nu}^\ast_{\ell'} \tilde{\nu}_{\ell'}$ in supersymmetry where the leptons $\ell$ and $\ell'$ can be either muon or electron type (see Fig.~\ref{fig:feyn2}). Even though the $\nu_1/\tilde{\nu}$ are not the LSP/LKP, they decay to neutrinos and the LSP, neither of which is visible in the detector.

\begin{figure}
\centering
\includegraphics[width=0.5\columnwidth]{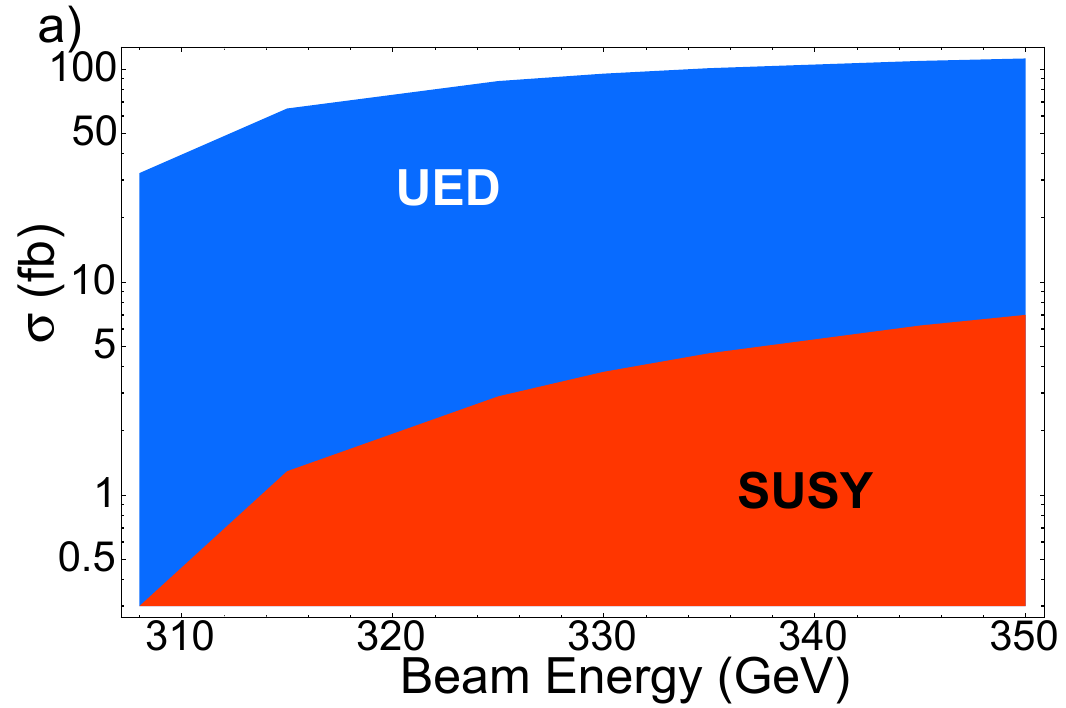}\includegraphics[width=0.5\columnwidth]{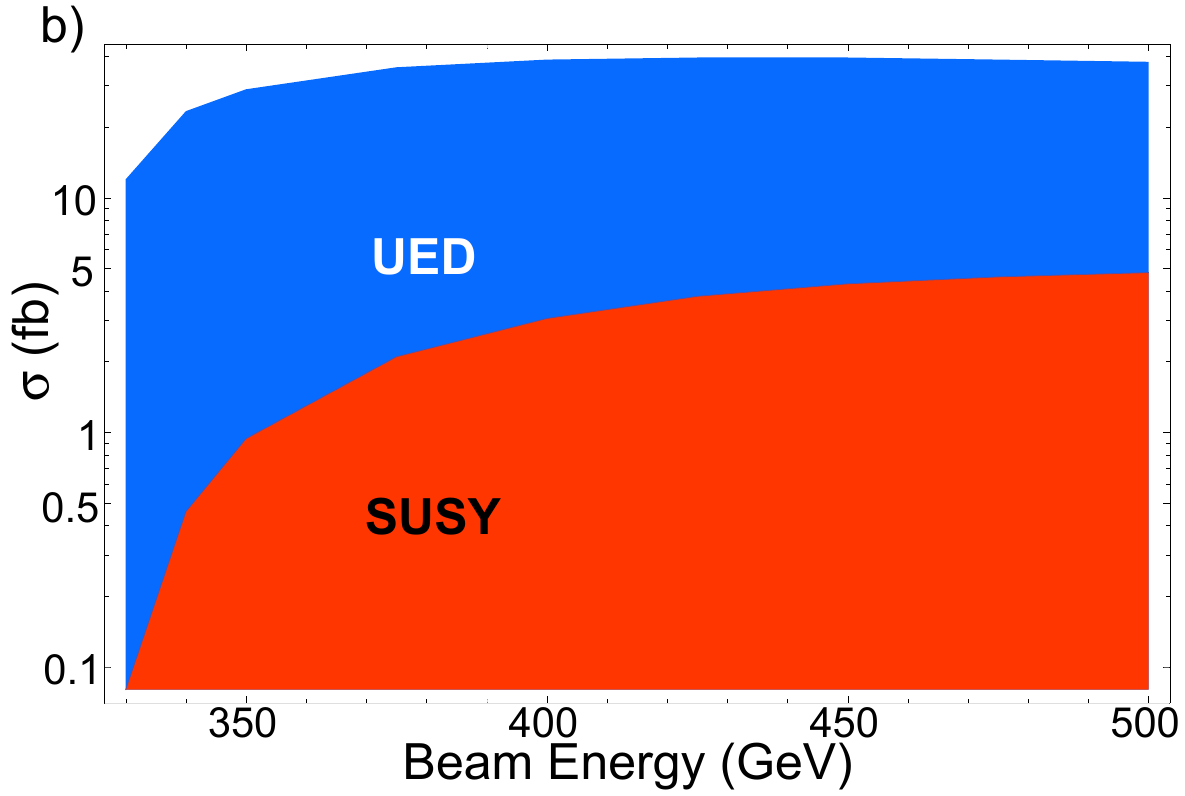}

\caption{Cross sections times branching ratios as a function of beam energy for the UED process $e_L^-e_L^+ \to W_1^-W_1^+\to \ell^- \ell'^+ \bar{\nu}_{1\ell}\nu_{1\ell'}$ and the SUSY process $e_L^-e_L^+ \to \tilde{\chi}_1^-\tilde{\chi}_1^+ \to \ell^- \ell'^+\tilde{\nu}^\ast_{\ell'} \tilde{\nu}_{\ell'}$. Figure a) uses the SPS 3 spectrum, while b) uses the MUED spectrum (see Table~\ref{tab:spectra}). \label{fig:sigmaW}}
\end{figure}

For these final states, the total cross sections times branching ratios as a function of energy are shown in Fig.~\ref{fig:sigmaW}. Once again, the supersymmetric cross section is considerably less than that in extra dimensions. Furthermore, the small mass splittings in the MUED spectrum lead to small cross sections compared to the SPS 3 case. Backgrounds include SM $W^\pm$ and $ZZ$ production, and model backgrounds from $\tilde{\chi}_2^0\tilde{\chi}_2^0/W_1^3W_1^3$, and $\tilde{\ell}^-\tilde{\ell}^+/\ell_1^-\ell_1^+$ pair production decaying to charged leptons and missing energy. However, we once again find that demanding successful reconstruction effectively cuts the background to negligible levels. In addition, we apply the $\eta \leq 2.5$ cuts on the charged leptons and missing momentum.

\begin{figure}
\centering
\includegraphics[width=\columnwidth]{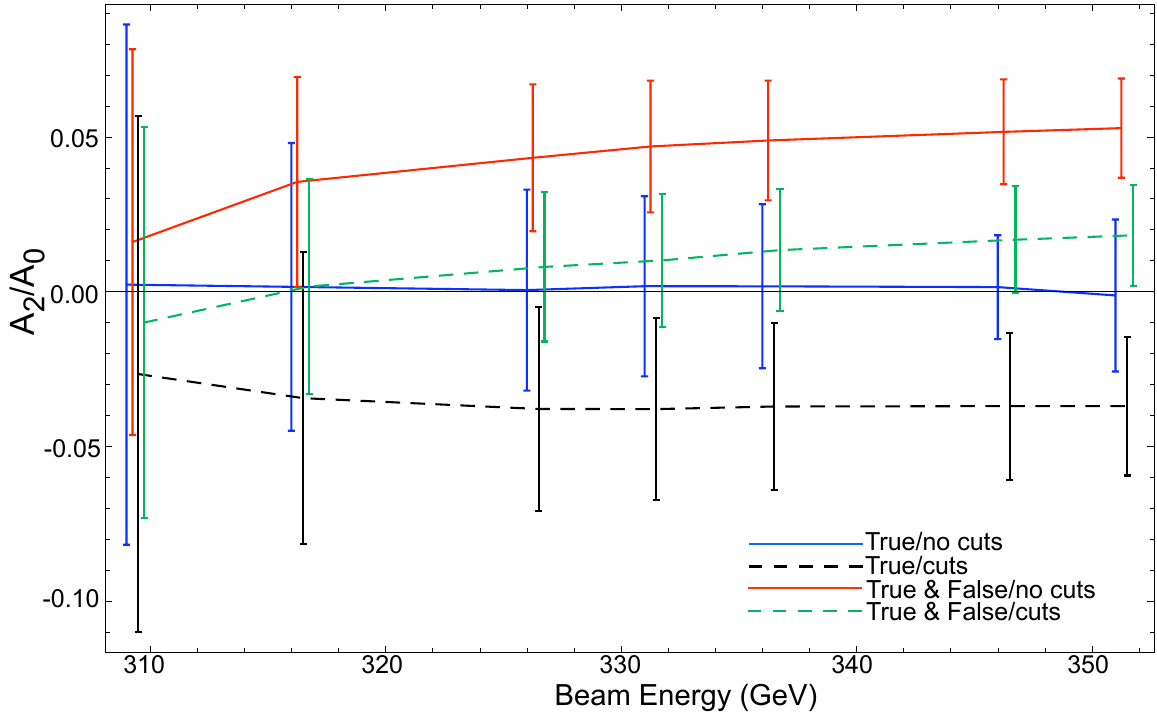}

\includegraphics[width=\columnwidth]{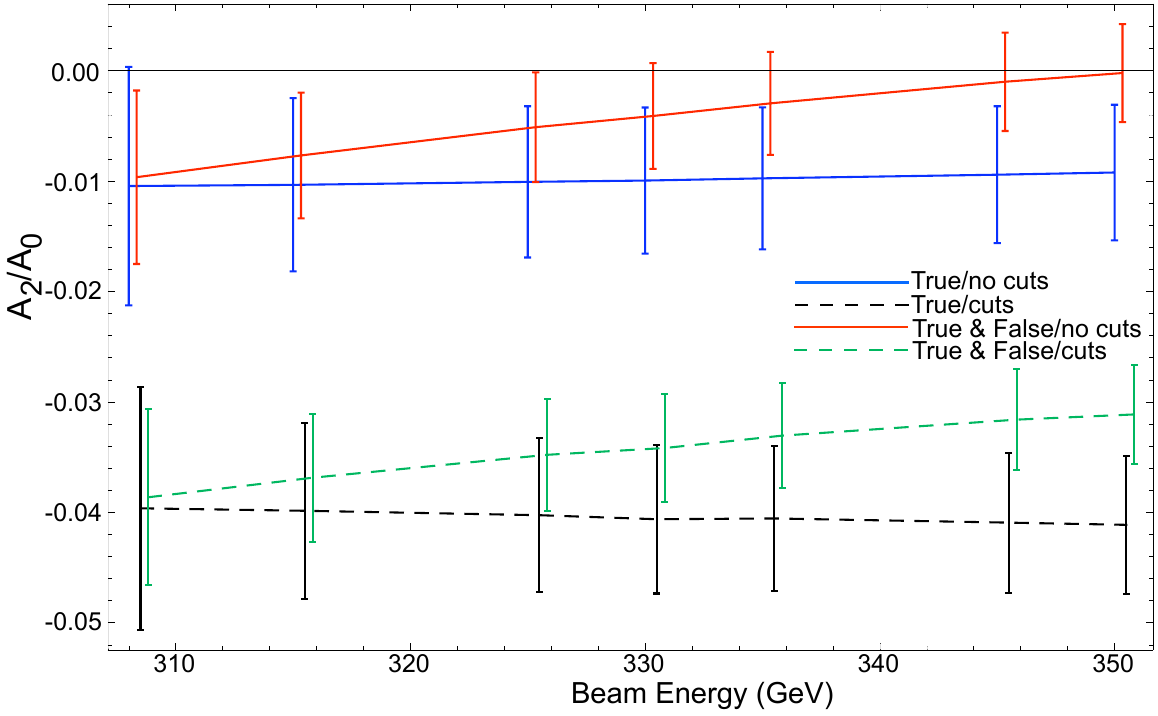}

\caption{Top: Ratio $A_2/A_0$ versus beam energy for the supersymmetric spinor production $e_L^-e_L^+ \to \tilde{\chi}_1^-\tilde{\chi}_1^+ \to \ell^- \ell'^+\tilde{\nu}^\ast_{\ell'} \tilde{\nu}_{\ell'}$ for the SPS3 spectrum. Bottom: Ratio $A_2/A_0$ for the UED vector boson production $e_L^-e_L^+ \to W_1^-W_1^+\to \ell^- \ell'^+ \bar{\nu}_{1\ell}\nu_{1\ell'}$ for the same spectrum.  Color coding as in Fig.~\ref{fig:a1a0}. Error bars correspond to 95\% exclusion region assuming $1~\mbox{ab}^{-1}$ luminosity. \label{fig:WSPS3}}
\end{figure}

\begin{figure}
\centering
\includegraphics[width=\columnwidth]{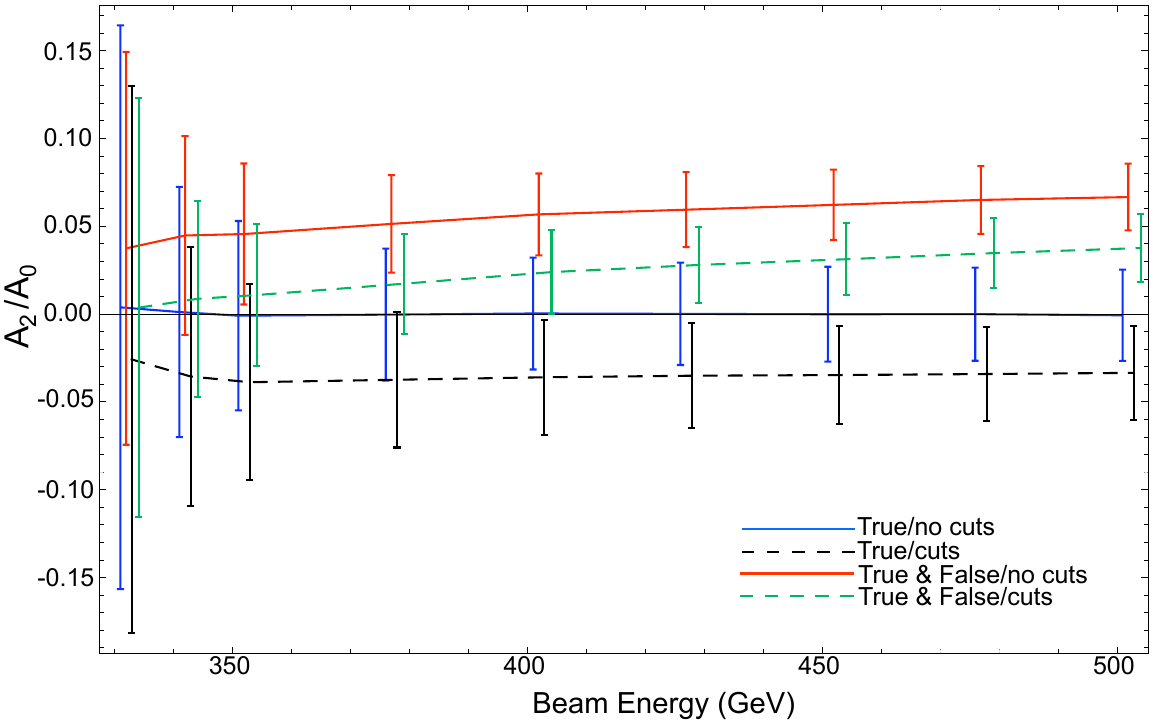}

\includegraphics[width=\columnwidth]{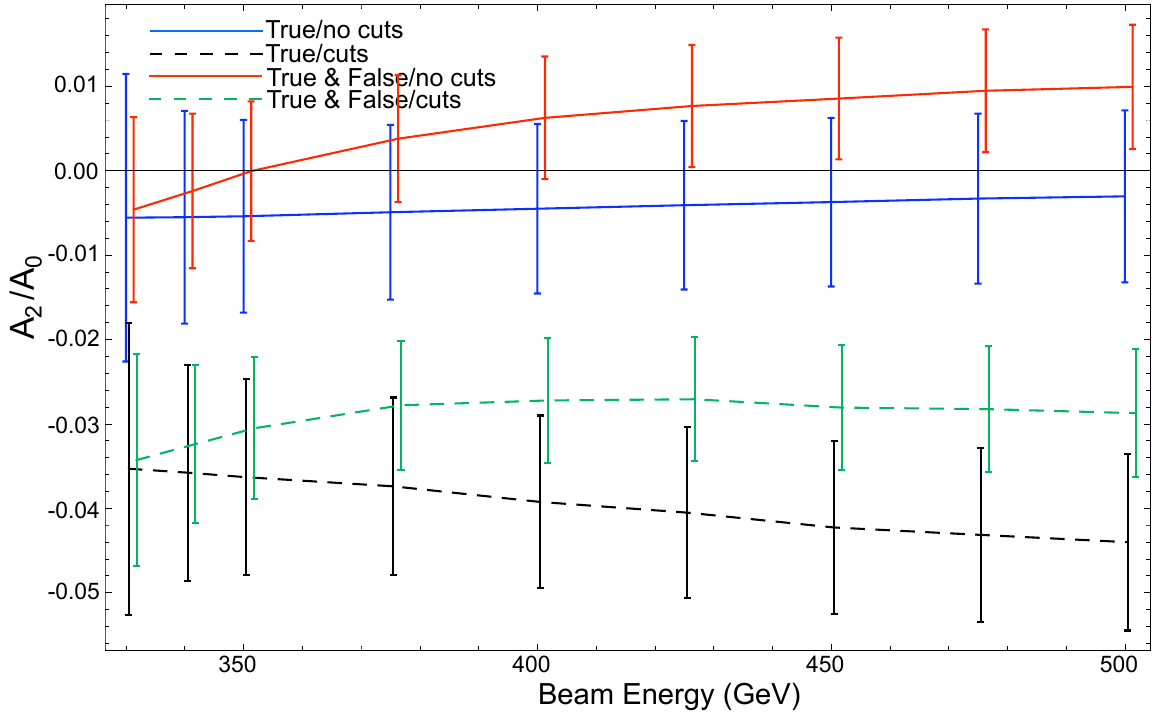}

\caption{Top: Ratio $A_2/A_0$ versus beam energy for the supersymmetric spinor production $e_L^-e_L^+ \to \tilde{\chi}_1^-\tilde{\chi}_1^+ \to \ell^- \ell'^+\tilde{\nu}^\ast_{\ell'} \tilde{\nu}_{\ell'}$ for the MUED spectrum (see Table~\ref{tab:spectra}). Bottom: Ratio $A_2/A_0$ for the UED vector boson production $e_L^-e_L^+ \to W_1^-W_1^+\to \ell^- \ell'^+ \bar{\nu}_{1\ell}\nu_{1\ell'}$ for the same spectrum.  Color coding as in Fig.~\ref{fig:a1a0}. Error bars correspond to 95\% exclusion region assuming $1~\mbox{ab}^{-1}$ luminosity. \label{fig:WMUED}}
\end{figure}

We perform fits to $A_0+A_1\cos\phi+A_2\cos2\phi$ as in section~\ref{sec:scalar} and consider the ratio $A_2/A_0$, using $1~\mbox{ab}^{-1}$ of integrated luminosity (due to the smaller cross sections). The results for the SPS3 spectrum are displayed in Fig.~\ref{fig:WSPS3}, and those of the MUED spectrum are shown in Fig.~\ref{fig:WMUED}. Note that the true solutions for the vector bosons consist of an approximately $1\%$ signal in the SPS3 spectrum and $\sim 0.5\%$ in MUED. In both spectra the true solution for spinors is consistent with zero. As with the production of $\mu_{1R}$ however, the presence of the false distribution introduces significant spurious values of $A_2/A_0$, dwarfing the true signal by a factor of $\sim 5$. 

In the SPS3 spectrum, even with $1~\mbox{ab}^{-1}$ the error bars on the true solution for the vector bosons barely exclude zero at $95\%$ confidence. For the MUED case, the situation is much worse, as a smaller signal is combined with cross sections suppressed by nearly an order of magnitude compared to those in the SPS3 case. Thus, statistics may be a limiting factor in measuring non-zero spins.

\begin{figure}
\centering
\includegraphics[width=\columnwidth]{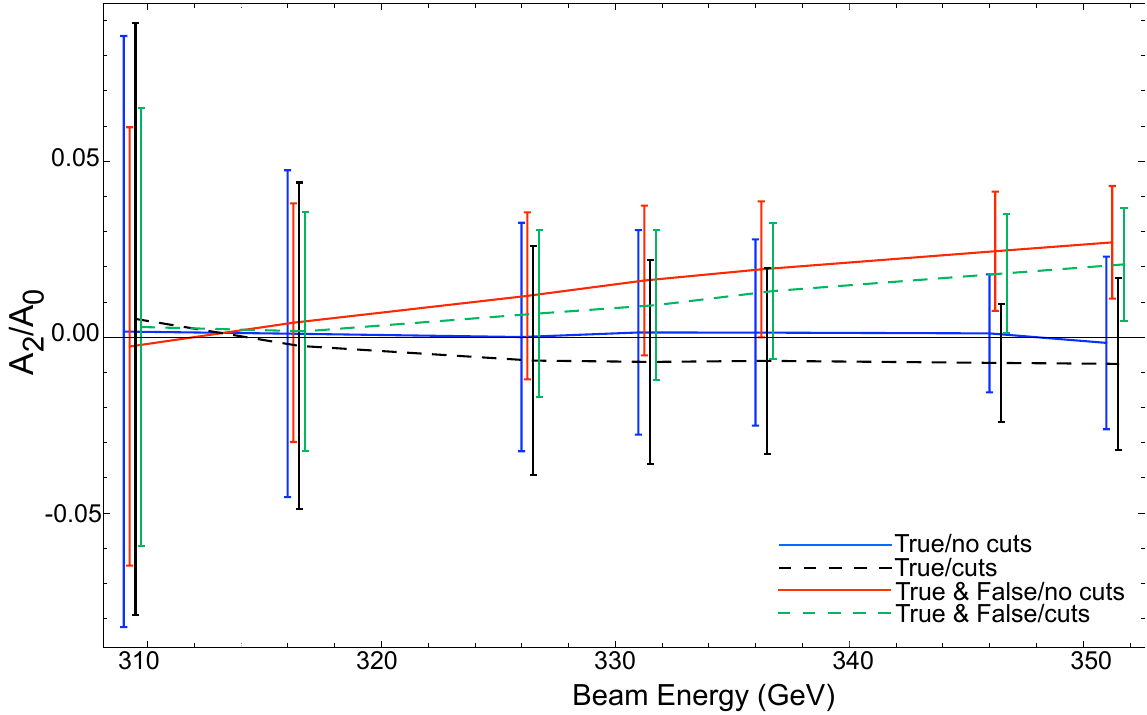}

\includegraphics[width=\columnwidth]{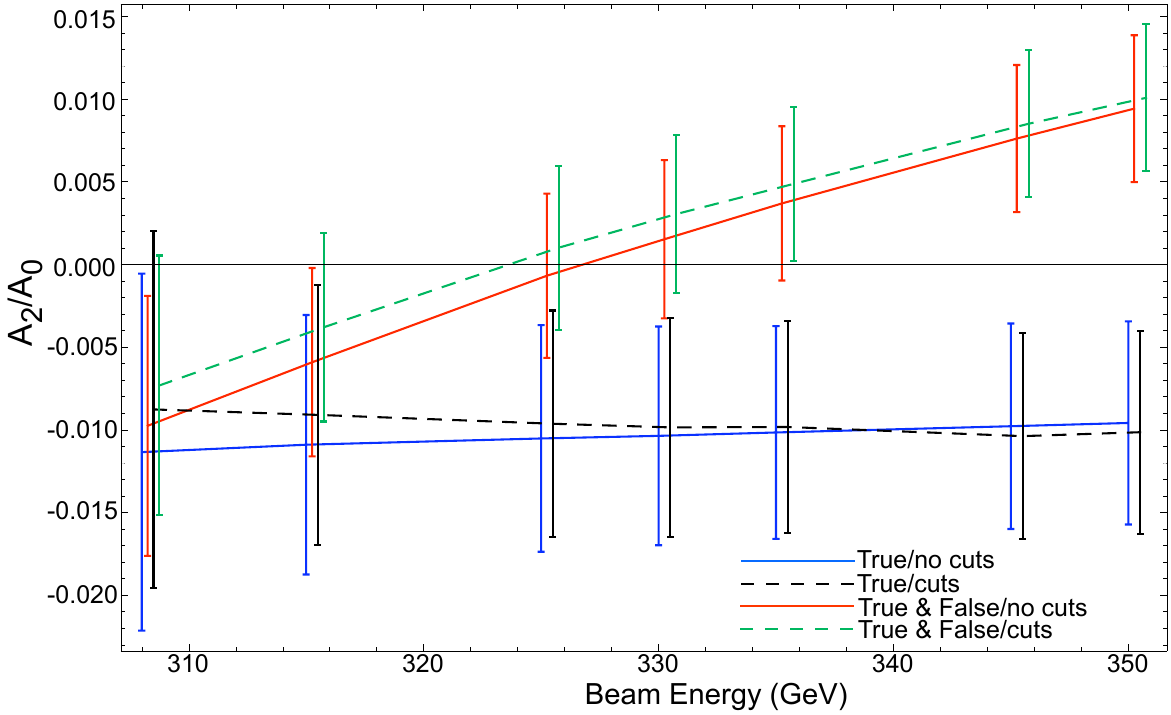}

\caption{Top: Ratio $A_2/A_0$ versus beam energy for the supersymmetric spinor production $e_L^-e_L^+ \to \tilde{\chi}_1^-\tilde{\chi}_1^+ \to \ell^- \ell'^+\tilde{\bar{\nu}}_{\ell'} \tilde{\nu}_{\ell'}$ for the SPS3 spectrum adjusted to account to detector and cut effects. Bottom: Ratio Adjusted values of $A_2/A_0$ for the UED vector boson production $e_L^-e_L^+ \to W_1^-W_1^+\to \ell^- \ell'^+ \bar{\nu}_{1\ell}\nu_{1\ell'}$ for the same spectrum.  Color coding as in Fig.~\ref{fig:a1a0}. Error bars correspond to 95\% exclusion region assuming $1~\mbox{ab}^{-1}$ luminosity. \label{fig:WSPS3ad}}
\end{figure}

We attempt to correct for the effects of cuts and false solutions by generating events which decay according to phase space. As the production angle may be measured up to the two-fold reconstruction ambiguity, we generate the particles with the correct $\theta$ distributions and flat $\theta_i$, $\phi_i$ distributions and run the resulting events through the reconstruction and detector simulator. The resulting values for $A_2/A_0$ are subtracted from those in Figs.~\ref{fig:WSPS3} and \ref{fig:WMUED} in an attempt to isolate the spin effects arising from quantum interference from the non-zero $A_2/A_0$ coming from cuts and the false solutions. The adjusted results are shown in Fig.~\ref{fig:WSPS3ad} for the SPS3 spectrum. Due to the small signal and poor statistics in the MUED spectrum, even the uncorrected signal in the true solution cannot be distinguished from zero, so we do not adjust for cuts or the false solutions.

Examining Fig.~\ref{fig:WSPS3ad}, we find that the flat distribution captures the effects of cuts on the ratio $A_2/A_0$ but does not correctly account for the false distributions. We do find that the false distributions do not contribute significantly to the ratio near threshold, as in the measurements of $A_1/A_0$. Once again, this behavior is not well understood and statistics in this regime are limited. It is conceivable that better results would be obtained by coupling a flat decay in $\phi_i$ with the {\it measured} distribution of $\theta_i$ to attempt to account for the false distribution. This matching has not  been performed as yet.

Thus, while the quantum interference measurement for spin-0 stands on solid ground, the situation for higher spins is less certain. Even neglecting the issue of false solutions, the vector boson ratio $A_2/A_0$ is on the order of $1\%$, and so requires significant statistics in order to distinguish from spinor decays. Furthermore, the false distribution introduces a spurious $A_2/A_0$ value which has not been fully understood by the authors and can dwarf the signal. Finally, the case of of the MUED spectrum demonstrates that, while the method of spin measurement is model independent, it is vulnerable to model-dependent effects such as total cross section, which control the statistical error of the fit. However, note that we could do much better statistically by adding hadronic final states for one of the $\tilde{\chi}^\pm_1/W_1$ while requiring leptonic final states for the other. We again would have two-fold ambiguity, but the rest of the measurement remains the same as long as we can measure the hadronic energies well enough. This may be possible by using the energy flow method that matches tracking and calorimeter information. 

\section{Conclusions \label{sec:conc}}

We have demonstrated that the quantum interference of multiple helicity states can provide a model independent method of spin measurements at the ILC. Specifically, with reasonable luminosities, scalar particles can be easily distinguished from spin-1/2 or higher possibilities in pair production followed by decays to visible leptons and missing energy. Determining whether a particle is spin-1/2 or spin-1 suffers from two major problems: the first is simply statistics: as the signal is on the order of 1\%, the requisite luminosity will be a stretch for the ILC, at least in the SUSY and UED models considered.

The second issue concerns the false solution to the reconstruction of the pair-produced particles' 4-momentum, and hence the derived values of the azimuthal angles $\phi_1$ and $\phi_2$. With 8 missing momentum components from the two weakly interacting particles escaping the detector, 4 measured total missing momenta, and 4 mass constraints the system can be solved only up to a two-fold ambiguity. While the $\cos\phi$ distribution is flat in the false solution, non-trivial dependences on $\cos2\phi$ develop. From explicit calculations, these dependences appear to be different for flat, spinor, and vector boson distributions, and so cannot be subtracted from the combined solutions without losing the desired model-independence.

It therefore behooves us to consider methods for full reconstruction of the event. If the decay proceeds by emitting several visible particles in a cascade of particles with known mass down to the LSP/LKP, then we may over-constrain the decay, allowing for full reconstruction. In particular, if the pair-produced particles decay to the LSP through an intermediate state, then there would be 6 mass constraints on the system. With only 8 unknown quantities and 4 measured values, the false solution is no longer present. Unfortunately, all such decays considered by the authors so far have too low a cross section to provide useful spin measurements.

However, such lengthy decay chains raises the possibility of applying this method to the LHC. At a hadron collider the center of mass energy and frame of reference are unknown for a particular parton-parton level event. Thus, only 2 measured quantities may be obtained in the event: the components of missing transverse momentum $\slashed{p}_T$. With a multi-step decay we obtain 6 mass constraints, combining these with the measured $\slashed{p}_T$ we can solve the system of 8 missing momentum components up to the two-fold ambiguity. Additionally, the reconstruction algorithm can be used in a modified form \cite{Cheng:2007xv} to measure the masses in the decay chain as a necessary preliminary step to determining the azimuthal angles. As the cross section for producing TeV-scale particles with color charge at the LHC is very large ({\it e.g.}~$\sim 1~\mbox{pb}$ for $\tilde{g}$ or $\tilde{q}$ pair production \cite{Paige:1997xb}) it seems likely that we may obtain enough statistics in such a case to at least measure the spin of scalar particles if not those of spin 1 or 1/2. This may possibly allow discrimination between the gluino and the KK gluon.

Note that the method we proposed can be used and tested already in the Tevatron top quark sample. The interference between the two helicity states of the top quark should give rise to $\cos\phi$ dependence. For the lepton$+$jet mode, one can fully reconstruct the event without a two-fold ambiguity, but this suffers from $W+$multi-jet background and the not-stellar jet energy resolution. The purely leptonic mode has two-fold ambiguity but less background and better momentum resolution. Run-II should already have enough statistics to attempt the study of azimuthal distributions, giving the first direct experimental hint on the spin $1/2$ nature of the top quark.

\appendix

\section{Reconstruction \label{sec:recon}}

The two charged leptons in the events shown in Fig.~\ref{fig:feyn1} have momenta $p_1$ and $p_2$ respectively. We define the perpendicular momentum in the event $\vec{p}_\perp = \vec{p}_1\times \vec{p}_2$. We refer to the pair produced unstable particles as $A$ (for $\mu^-_{1R}$ or $\tilde{\mu}_R^-$) and $B$ ($\mu^+_{1R}$ or $\tilde{\mu}_R^+$). The missing 4-momentum from the decay of $A$ is $\slashed{p}_1$, while $\slashed{p}_2$ is the missing momentum from the decay of $B$. Both the particles escaping the detector have mass $m$, which is assumed to be known.

Since the pair produced particles $A$ and $B$ (with mass $M$) are back to back, it suffices to solve for $p_A$, as $\vec{p}_A = -\vec{p}_B$. The final state leptons are effectively massless, so $p_1^2 = p_2^2 =0$. For the massive particles, we have
\begin{eqnarray}
p_A^2 & = & p_B^2 = M^2 \label{eq:pa} \\
\slashed{p}_1^2 & = & \slashed{p}_2^2 = m^2 \label{eq:pslash}
\end{eqnarray}
Finally, since $p_A$ ($p_B$) decays into $\slashed{p}_1$ ($\slashed{p}_2$) and $p_1$ ($p_2$),
\begin{eqnarray}
\slashed{p}_1 & = & p_A - p_1 \nonumber \\
\slashed{p}_2 & = & p_B -p_2 \label{eq:momcon}
\end{eqnarray}

At the ILC, the energy of the beams $E$ is known, and for pair production the total energy in the event must be split equally, so $p_A^0= p_B^0=E$. Therefore, using Eqs.~(\ref{eq:pa}), (\ref{eq:pslash}) and (\ref{eq:momcon}) we may define the following variables
\begin{eqnarray}
c_1 & \equiv & \vec{p}_A \cdot \vec{p}_1 = \frac{1}{2}(m^2-M^2+2Ep_1^0) \label{eq:c1} \\
c_2 & \equiv & \vec{p}_A \cdot \vec{p}_2 = -\frac{1}{2}(m^2-M^2+2Ep_2^0) \label{eq:c2} \\
b_2 & \equiv & \vec{p}_A\cdot\vec{p}_A = E^2 - M^2 \label{eq:b2} \\
a_{ij} & \equiv & \vec{p}_i \cdot \vec{p}_j~(i,j = 1,2) \label{eq:aij}
\end{eqnarray}
We can write the momentum $\vec{p}_A$ as
\begin{equation}
\vec{p}_A = t_1 \vec{p}_1+t_2\vec{p}_2 + y \vec{p}_\perp. \label{eq:parecon}
\end{equation}
Using this definition in Eq.~(\ref{eq:c1}) and (\ref{eq:c2}), we find
\begin{eqnarray}
c_1 & = &  t_1 a_{11}+t_2a_{12} \nonumber \\
c_2 & = & t_1 a_{12} +t_2a_{22} \nonumber \\
t_1 & = & \frac{a_{22}c_1-a_{12}c_2}{a_{11}a_{22}-a_{12}^2} \label{eq:t1} \\
t_2 & = & \frac{a_{11}c_2-a_{12}c_1}{a_{11}a_{22}-a_{12}^2} \label{eq:t2}
\end{eqnarray}
Finally, using Eqs.~(\ref{eq:b2}), (\ref{eq:t1}) and (\ref{eq:t2})
\begin{eqnarray}
b_2 & = & (t_1^2a_{11}+2t_1t_2 a_{12}+t_2^2 a_{22})+y|\vec{p}_{\perp}|^2 \nonumber \\
y & = & \pm\sqrt{\frac{b_2 - (t_1^2a_{11}+2t_1t_2 a_{12}+t_2^2 a_{22})}{|\vec{p}_{\perp}|^2}} \label{eq:y}
\end{eqnarray}
The $\pm$ sign in this last equation is the two-fold ambiguity in the reconstruction.

\section{Amplitudes \label{sec:cross}}

The matrix elements for right-handed smuon pair-production from polarized $e^-e^+$ beams are 
\begin{eqnarray}
\lefteqn{{\cal M}(e_{L}^-e_{R}^+ \to \tilde{\mu}_R^-\tilde{\mu}_R^+)  =  (-ie^2) \sqrt{1-\frac{4m_{\tilde{\mu}}^2}{s^2}} \sin\theta \times} \hspace{5em}\nonumber \\
 & & \left(1 +\frac{s (-1/2+s_W^2)}{c_W^2 (s-4m_Z^2)}\right) \label{eq:esmu} \\
 \lefteqn{{\cal M}(e_{R}^-e_{L}^+ \to \tilde{\mu}_R^-\tilde{\mu}_R^+)  = (-ie^2) \sqrt{1-\frac{4m_{\tilde{\mu}}^2}{s^2}} \sin\theta\times } \hspace{5em}\nonumber \\
 & & \left(1 +\frac{s_W^2 s}{c_W^2 (s-4m_Z^2)} \right)  \nonumber
\end{eqnarray}
Here, $\sqrt{s}$ is the center of mass energy and $m_{\tilde{\mu}}$ is the mass of right-handed smuon. The angle $\theta$ is defined as in Fig.~\ref{fig:angles}.

The decaying $\tilde{\mu}^\pm$ goes to $\mu_R^\pm$ and a right-handed $\tilde{\chi}_1^0$. We make the approximation that the neutralino is primarily bino, and so the decay matrix element is
\begin{equation}
{\cal M}(\tilde{\mu}^\pm \to \mu^\pm \tilde{\chi}_1^0) = -\sqrt{2}g'\sqrt{m_{\tilde{\mu}}^2-m_{\tilde{\chi}}^2} \label{eq:smudecay}
\end{equation}
Here $g'$ is the hypercharge gauge coupling $g' = \frac{e}{\cos\theta_w}$. Making the narrow width approximation, the cross section for the four-body final state is simply the incoherent sum over initial helicities
\begin{eqnarray}
d\sigma & = & \frac{d \Phi_4 }{4}\sum_{L,R} \left|{\cal M}(e_{L/R}^-e_{R/L }^+\to \tilde{\mu}^-\tilde{\mu}^+) {\cal M}(\tilde{\mu}_R^-\to \mu^-_R\tilde{\chi}_1^0)\times\right. \nonumber \\
& &\left. {\cal M}(\tilde{\mu}_R^+\to \mu^+_R\tilde{\chi}_1^0) \right|^2 2\pi  \delta(s_{\mu^+ \tilde{\chi}_1^0}-m_{\tilde{\mu}}^2)\times \nonumber \\
 & &  2\pi \delta(s_{\mu^- \tilde{\chi}_1^0}-m_{\tilde{\mu}}^2) \frac{1}{(2m_{\tilde{\mu}}\Gamma)^2}  \label{eq:Asmu}
\end{eqnarray}
where $\Gamma$ is the total width of the $\tilde{\mu}_{1R}$. Note the lack of dependence on $\phi_1$ and $\phi_2$, in accordance with Eq.~(\ref{eq:scalar}).

Pair production for right-handed mu-1 requires four helicity combinations for the $\mu_{1R}$s. Recall that KK states of the chiral muons are massive particles, as such they can have either helicity. Thus, the production matrix elements are
\begin{eqnarray}
{\cal M}(e_L^-e_R^+\to \mu_{1R}^-(\downarrow)\mu_{1R}^+(\uparrow)) & = & (ie^2)(1-\cos \theta) \times \nonumber \\
 & & \left(1+\frac{s(-1/2+s_W^2)}{c_W^2(s-m_Z^2)} \right) \nonumber \\
{\cal M}(e_R^-e_L^+\to \mu_{1R}^-(\downarrow)\mu_{1R}^+(\uparrow)) & = &  (-ie^2)(1+\cos \theta) \times \nonumber \\
 & &  \left(1+\frac{s_W^2 s}{c_W^2(s-m_Z^2)} \right) \nonumber \\
 {\cal M}(e_{L}^-e_{R}^+\to \mu^-_{1R}(\downarrow)\mu^+_{1R}(\downarrow)) & = & (-ie^2)\frac{2m_{\mu_1}}{\sqrt{s}} \sin\theta \nonumber \\
 & & \left(1+\frac{s(-1/2+s_W^2)}{c_W^2(s-m_Z^2)} \right) \nonumber \\
  {\cal M}(e_{R}^-e_{L}^+\to \mu^-_{1R}(\downarrow)\mu^+_{1R}(\downarrow)) & = & (-ie^2)\frac{2m_{\mu_1}}{\sqrt{s}} \sin\theta \nonumber \\
 & &  \left(1+\frac{s_W^2 s}{c_W^2(s-m_Z^2)} \right) \nonumber \\
{\cal M}(e_L^-e_R^+\to \mu^-_{1R}(\uparrow)\mu^+_{1R}(\downarrow)) & = & (-ie^2)(1+\cos \theta) \times \nonumber \\
 & & \left(1+\frac{s(-1/2+s_W^2)}{c_W^2(s-m_Z^2)} \right) \nonumber \\
 {\cal M}(e_R^-e_L^+\to \mu^-_{1R}(\uparrow)\mu^+_{1R}(\downarrow)) & = & (ie^2)(1-\cos \theta) \times \nonumber \\
 & & \left(1+\frac{s_W^2 s}{c_W^2(s-m_Z^2)} \right) \nonumber \\
{\cal M}(e_{L}^-e_{R}^+\to \mu^-_{1R}(\uparrow)\mu^+_{1R}(\uparrow)) & = & (ie^2)\frac{2m_{\mu_1}}{\sqrt{s}} \sin\theta \nonumber \\
 & & \left(1+\frac{s(-1/2+s_W^2)}{c_W^2(s-m_Z^2)} \right) \nonumber \\
   {\cal M}(e_{R}^-e_{L}^+\to \mu^-_{1R}(\uparrow)\mu^+_{1R}(\uparrow)) & = & (ie^2)\frac{2m_{\mu_1}}{\sqrt{s}} \sin\theta \nonumber \\
 & & \left(1+\frac{s_W^2 s}{c_W^2(s-m_Z^2)} \right) \nonumber \\
 \label{eq:emu}
\end{eqnarray}
Here, $\uparrow$ corresponds to right-handed helicity, while $\downarrow$ is left-handed.

In the rest frame of the decaying $\mu_{1R}$, there are two possible helicities ($\uparrow$ and $\downarrow$) decaying to right-handed muons and three possible polarization vectors for the $B_1$ ($\epsilon_\lambda$, $\lambda = \pm 1, 0$). For the decay of the $\mu_{1R}^-$, the matrix elements are
\begin{eqnarray}
{\cal M}(\mu_{1R}^-(\uparrow) \to \mu_R^- B_1(-1)) & = & 0 \nonumber \\
{\cal M}(\mu_{1R}^-(\uparrow) \to \mu_R^- B_1(0)) & = & g' \frac{m_{\mu_1}}{m_{B_1}}\sqrt{m_{\mu_1}^2-m_{B_1}^2} \times \nonumber \\
 & & e^{+i\phi_1/2} \cos\frac{\theta_1}{2} \nonumber \\
{\cal M}(\mu_{1R}^-(\uparrow) \to \mu_R^- B_1(+1)) & = & -\sqrt{2}g' \sqrt{m_{\mu_1}^2-m_{B_1}^2} \times \nonumber \\
 & & e^{+i\phi_1/2} \sin\frac{\theta_1}{2} \nonumber \\
{\cal M}(\mu_{1R}^-(\downarrow) \to \mu_R^- B_1(-1)) & = & 0 \nonumber \\
{\cal M}(\mu_{1R}^-(\downarrow) \to \mu_R^- B_1(0)) & = & g' \frac{m_{\mu_1}}{m_{B_1}}\sqrt{m_{\mu_1}^2-m_{B_1}^2}\times \nonumber \\
 & & e^{-i\phi_1/2} \sin\frac{\theta_1}{2} \nonumber\\
 {\cal M}(\mu_{1R}^-(\downarrow) \to \mu_R^- B_1(+1)) & = & \sqrt{2}g' \sqrt{m_{\mu_1}^2-m_{B_1}^2}\times \nonumber \\
 & & e^{-i\phi_1/2} \cos\frac{\theta_1}{2} \label{eq:mu1decay}
\end{eqnarray}
We see here the dependence on the helicity of the $\mu_{1R}$ as in Eq.~(\ref{eq:spinor}). Similar equations hold for the decay of $\mu_{1R}^+$, with $\phi_1 \to \phi_2$ and $\theta_1\to\theta_2$.

The total cross section for the event is the coherent sum over $\mu_{1R}$ helicities and the incoherent sum over the helicities $h$ of the electrons and polarizations $\lambda$ of the KK photons:
\begin{eqnarray}
d\sigma & = & \frac{d \Phi_4 }{4}\sum_{L,R,\lambda\lambda'} \left| \sum_{h h'}{\cal M}(e_{L/R}^-e_{R/L}^+ \to \mu_{1R}^-(h)\mu^+_{1R}(h'))  \right. \nonumber  \\
& & \left. {\cal M}(\mu_{1R}^-(h) \to \mu^-_R B_1(\lambda)){\cal M}(\mu_{1R}^+(h') \to \mu^+_R B_1(\lambda'))\right|^2 \nonumber \\
 & & 2\pi \delta(s_{\mu^+ B_1}-m_{\mu_1}^2) 2\pi \delta(s_{\mu^+ B_1}-m_{\mu_1}^2) \frac{1}{(2m_{\mu_1}\Gamma)^2} \label{eq:Amu1}
\end{eqnarray}
Once again, $\Gamma$ is the total width of $\mu_{1R}$ and there is an implied momentum conserving $\delta$ function.

\begin{acknowledgments}
  This work was supported in part by the U.S. DOE under Contract
  DE-AC03-76SF00098, and in part by the NSF under grant PHY-04-57315.
\end{acknowledgments}

\end{document}